\documentclass{emulateapj}
\usepackage{graphicx}
\usepackage{amssymb}
\usepackage{amsmath}
\usepackage{subfigure}
\usepackage{multirow}
\usepackage{threeparttable}
\usepackage{array}
\usepackage{natbib}
\usepackage{xcolor}
\usepackage{subfigure}
\usepackage{hyperref}

\hypersetup{
    colorlinks=true,
    linkcolor=red,
    filecolor=magenta,
    urlcolor=cyan,
}

\def\kms{~\rm km~s^{-1}}
\def\cmsq{\rm cm^{-2}}
\def\cc{\rm cm^{-3}}

\defcitealias{Savage:2009aa}{SW09}
\defcitealias{Savage:2003aa}{Savage03}
\defcitealias{Zheng:2019aa}{Zheng19}

\begin{document}
\title{The warm Gaseous Disk and the Anisotropic Circumgalactic Medium of the Milky Way}
\author{Zhijie Qu$^1$ and Joel N. Bregman$^1$}
\affil{$^1$ Department of Astronomy, University of Michigan, Ann Arbor, MI 48109, USA}
\email{quzhijie@umich.edu}

\begin{abstract}
The warm ($\log~T\approx5$) gas is an important gaseous component in the galaxy baryonic cycle.
We built a 2-dimension disk-CGM model to study the warm gas distribution of the Milky Way (MW) using the absorption line surveys of \ion{Si}{4} and \ion{O}{6}.
In this model, the disk component of both ions has the same density profile ($n(r, z)=n_0\exp(-|z|/z_0)\exp(-r/r_0)$) with a scale height of $z_0=2.6\pm0.4\rm~kpc$ and a scale length of $r_0=6.1\pm1.2\rm~kpc$.
For this disk component, we calculate the warm gas mass of $\log(M/M_\odot)=(7.6\pm0.2)-\log\frac{Z}{Z_\odot}$.
The similar disk density profiles and total masses of \ion{Si}{4} and \ion{O}{6}-bearing gas set constraints on the ionization mechanisms.
We suggest that the warm gas disk might be dominated by the Galactic fountain mechanism, which ejects and recycles gas to set both the scale height and the scale length of the warm gas disk.
The CGM component in our model has a dependence on Galactic latitude with a higher column density along the direction perpendicular to the Galactic plane ($b=90^\circ$) than the column density along the radial direction ($b=0^\circ$).
The column density difference between these two directions is $0.82\pm0.32\rm~dex$ at $6.3\sigma$ for both ions.
This difference may be due to the enrichment of Galactic feedback to the entire CGM, or an additional interaction layer between the warm gas disk and the CGM; existing data cannot distinguish between these two scenarios.
If this higher column density at $b=90^\circ$ is for the entire CGM, the total warm CGM mass is $\log(M/M_\odot)\approx(9.5-9.8)-\log\frac{Z}{0.5Z_\odot}$ within the MW virial radius of $250\rm~kpc$.
\end{abstract}
\maketitle

\section{Introduction}
The gaseous baryons in galaxies can be found in both gaseous disks (the interstellar medium; \citealt{Dickey:1990aa}) and gaseous halos (the circumgalactic medium; CGM; \citealt{Putman:2012aa, Tumlinson:2017aa}).
The gaseous disks are roughly cospatial with the galaxy stellar disks, whereas the CGM are surrounding the stellar disks.
\ion{H}{1} 21 cm line surveys or ultraviolet (UV) observations reveal that the gaseous disks can extend beyond the stellar disks (up to $30-50$ kpc along the major axis) with masses from $10^7~M_\odot$ to $\lesssim 10^{10}~M_\odot$ for low-redshift galaxies ($z \lesssim 0.2$; \citealt{Oosterloo:2007aa, Bregman:2018aa}).
With statistical assembly of Quasar-galaxy pairs in UV absorption lines, the CGM can be detected out to large radii ($> 150\rm~ kpc$) and contribute a large amount of baryonic materials ($\gtrsim 10^{10}~M_\odot$) for galaxies at $z\lesssim 0.5$ \citep{Stocke:2013aa, Werk:2014aa, Lehner:2015aa, Keeney:2017aa}.

The existence of these gaseous components is important for galaxy evolution by mediating both accretion and feedback processes \citep{Mo:2010aa}.
For $L \gtrsim L^*$ galaxies, the CGM is normally volume-filled by warm-hot gas ($\log T \approx 5-7$) together with discrete cool gas clouds ($\log T \approx 4-5$), which is mainly shock-heated by gas accreted from the intergalactic medium (IGM) and altered by feedback processes from host galaxies \citep{Cen:2006aa}.
The existence of the CGM could prevent direct accretion from the IGM; instead, it provides cooling materials from itself by thermal or gravitational instabilities \citep{Keres:2009aa}.
These cooling flows from the CGM supplement the gaseous disk, and sustain the star formation activities in the stellar disk \citep{Lehner:2011aa, Li:2014aa, Borthakur:2015aa, Qu:2018aa}.
In turn, the stellar and active galactic nuclei (AGN) feedback could enrich the CGM by ejecting gas, energy, and metals \citep{Borthakur:2013aa, Fielding:2017ab, Oppenheimer:2018aa}.

{The warm ($\log~T\approx5$) gas is of special importance, since it is at transitional temperatures (the peak of the radiative cooling curve).
In this temperature range, the gas could be the interaction layer between the cool gas and the warm-hot gas.
These interactions are mostly associated with galactic feedback (e.g., outflows) or gas accretion onto the disk (e.g., accretion shocks).
Therefore, the distribution of the warm gas is crucial to investigate these fundamental processes.
Observationally, the warm gas distribution could be divided into two components: the warm gas disk and the warm CGM.
The warm gas disk has been studied for both other galaxies \citep{Boettcher:2016aa, Zheng:2017aa, Qu:2019aa} and the Milky Way (MW; \citealt{Howk:2002aa, Finkbeiner:2003aa, Savage:2003aa, Wakker:2012aa}).}
For other galaxies, the warm gas disk is detected in various observations.
Direct imaging on nearby edge-on galaxies has detected the warm gas disk at radii of $\approx 1- 10\rm~ kpc$ in X-ray band, UV band, or nebula emission lines \citep{Rand:2008aa, Li:2013aa, Boettcher:2016aa, Hodges-Kluck:2016ab}.
However, for the warm CGM, these observations are limited at larger radii ($>50$ kpc) due to the current instrument limitations and the low surface brightness of the diffuse ionized gas.
An alternative is to use UV absorption lines against the continua of background AGN/stellar objects to detect gas with column densities as low as $\approx 10^{12}~\rm cm^{-2}$ (\citealt{Tumlinson:2017aa} and reference therein).
The warm gas is traced by intermediate-high ionization state UV ions, such as \ion{Al}{3}, \ion{Si}{4}, \ion{C}{4}, and \ion{O}{6} \citep{Stocke:2013aa, Werk:2014aa, Johnson:2015aa, Lehner:2015aa, Zheng:2017aa, Qu:2019aa}.
However, extragalactic absorption-line studies are all limited by the sample of available sightlines for individual galaxies.

The MW is a unique target to study the warm gas distribution with hundreds of sightlines over the sky mapping both the disk and the CGM \citep{Savage:1997aa, Howk:2002aa, Wakker:2003aa, Savage:2009aa, Lehner:2011aa, Lehner:2012aa, Wakker:2012aa, Fox:2014aa, Fox:2015aa, Bordoloi:2017aa, Karim:2018aa}.
{Previous studies indicated that the warm gas could be discrete kpc-size clouds, which can be detected as intermediate-velocity clouds (IVCs) or high-velocity clouds (HVCs; \citealt{Sembach:2003aa, Wakker:2003aa, Fox:2004aa, Shull:2009aa, Wakker:2012aa, Werk:2019aa}).
At large scales, the warm gas distribution in the MW was modeled as a plane-parallel slab with only one dimensional (1-D) variation over the vertical direction (perpendicular to the disk) as an exponential function of $n(z) = n_0 \exp(-|z|/z_{0})$, where $z_0$ is the scale height (\citealt{Savage:2009aa} and references therein; hereafter \citetalias{Savage:2009aa}).}
The scale height of the warm gas disk is measured using column densities against UV-bright stars at different $z$-heights and AGNs.
In this model, the stellar sightlines are used to estimate the ion density at the mid-plane of the disk.
Combining with the sightlines toward AGNs (determining the maximum projected column density), the scale heights of the plane-parallel slab model are obtained for various ions, where no CGM component is considered.
However, more recent observations reveal the CGM also contributes to the column densities of the intermediate-high ionization state ion measured in the AGN sightlines \citep{Werk:2014aa, Johnson:2015aa, Lehner:2015aa, Stocke:2017aa, Zahedy:2019aa}.
As pointed out by  \citet[][hereafter \citetalias{Zheng:2019aa}]{Zheng:2019aa}, the MW CGM contributes to a significant amount of column density to the \ion{Si}{4} absorption lines measured with 132 AGN sightlines obtained by the Cosmic Origins Spectrograph (COS; \citealt{Green:2012aa}) on the {\it Hubble Space Telescope} ({\it HST}).
They demonstrated that the warm gas in the MW, as observed with all-sky AGN sightlines, should be modeled with a two-component model (i.e., a disk component and a CGM component). 
In their two-component model, the disk component follows the 1-D plane-parallel slab model as \citetalias{Savage:2009aa} adopted, and the halo component is modeled as a uniform global background. 

This paper is built upon the two-component model of \citetalias{Zheng:2019aa} to develop a disk-CGM model that accounts for both the radial and vertical density profiles of the disk.
This model is applied to \ion{Si}{4} and \ion{O}{6}, which are typical ions tracing the transitional temperature gas \citep{Savage:1997aa, Wakker:2012aa}.
In Section 2, we summarize the data used in this study, which includes column density measurements from MW stellar sightlines by \citetalias{Savage:2009aa}, and from all-sky AGN sightlines by \citet[][hereafter \citetalias{Savage:2003aa}]{Savage:2003aa} and \citetalias{Zheng:2019aa}. 
We introduce our model in Section 3, and show that the inclusion of the disk radial profile can alleviate the disagreement between the plane-parallel slab model based on the stellar sample \citepalias{Savage:2009aa} and the two-component model based on the AGN sample \citepalias{Zheng:2019aa}.
The anisotropic CGM model is also introduced in Section 3, showing that the MW is likely to have a warm CGM with anisotropic column density distribution.
In Section 4, we discuss the implication of this work on the warm gas disk (Section 4.2), the warm CGM of the MW (Section 4.3), and the north-south asymmetry of the MW warm gas absorption features (Section 4.4).
The key results are summarized in Section 5.

\section{Data}
At the temperature of $\approx 10^5\rm~K$, the transitional gas can be traced by intermediate-high ionization state ions, such as \ion{Si}{4} with ionization potential of 33.5 - 45.1 eV, \ion{C}{4} (47.9 - 64.5 eV), \ion{N}{5} (77.5 - 97.9 eV), or \ion{O}{6} (113.9 - 138.1 eV).
These ions are detectable in absorption against the continua of background UV-bright stars or AGNs.
For observations of the MW warm gas disk and CGM, the stellar sightlines are normally at low Galactic laitutdes ($|b|\lesssim5^\circ$; \citetalias{Savage:2009aa}), whereas the AGNs are at high Galactic latitudes ($|b|\gtrsim30^\circ$; \citetalias{Savage:2003aa, Zheng:2019aa}).
The stellar sightlines are employed to measure the mid-plane density of the disk, while the AGN sightlines can trace the large-scale variation of the disk (e.g., scale height of the disk) and the CGM.
Therefore, these two samples are equally important to constrain both the gaseous disk shape and the CGM contribution.
We only consider the ions that have high $S/N\gtrsim15$ samples for both disk stars and AGNs of sample sizes $\mathcal{N} \gtrsim 100$.

Based on this criterion, \ion{Si}{4} and \ion{O}{6} are the two ions, which have both stellar sightlines from \citetalias{Savage:2009aa} and AGN sightlines from \citetalias{Savage:2003aa} and \citetalias{Zheng:2019aa}.
{In this study, we do not consider \ion{C}{4}, since the current largest AGN sample ($\mathcal{N} \approx 30-40$) from \citet{Wakker:2012aa} does not have sufficient sightlines to obtain a good fitting result.}
All of the used three samples \citepalias{Savage:2003aa, Savage:2009aa, Zheng:2019aa} have similar velocity ranges ($|v|\lesssim 100\kms$) for the measurement of the column density, so we only study the low-intermediate velocity gas of the MW (without HVCs $|v|\gtrsim 100\kms$).
\citet{Zheng:2015aa} showed that a significant amount of the CGM is at low-intermediate velocities using a MW-mass hydrodynamic simulation \citep{Joung:2012aa}, so we expect that we could detect both the warm gas disk and CGM using these samples.

For the stellar sample, \citetalias{Savage:2009aa} summarized the {\it FUSE}, {\it IUE}, and {\it Copernicus} sightlines toward the 109 MW stars, 25 AGNs, and 6 LMC/SMC stars with good measurements ($\sigma_N < 0.4$ dex) or limits of \ion{Si}{4} and \ion{O}{6}.
This sample is mainly based on the Galactic \ion{O}{6} surveys, such as \citet{Bowen:2008aa} for low-latitude disk stars and \citet{Zsargo:2003aa} for halo stars.
Compared to the \citet{Bowen:2008aa} sample, \citetalias{Savage:2009aa} excluded sightlines with large uncertainties of \ion{O}{6} and other transitional ions (e.g., \ion{C}{4}, \ion{Si}{4}; \citealt{Savage:2001aa}). 
The excluded sightlines are mainly stellar sightlines within 1 kpc. 
Since \citetalias{Savage:2009aa}  also included sightlines toward halo stars at $|z|>1$ kpc, this sample is better to constrain the scale height of the disk component.
We exclude sightlines that might be contaminated by foreground \ion{H}{2} regions as marked out by \citetalias{Savage:2009aa}.
Besides the Galactic stellar sample, \citetalias{Savage:2009aa} also included 6 stars in LMC/SMC, which are not used in our analyses.
This is because one needs to assume the radial density profile of the MW CGM to model the ion column densities from stars at the distance of $50-60\rm~kpc$, which is highly uncertain.
Therefore, we do not implement this variation in our model, and omit the sightlines toward LMC/SMC stars.
The final stellar sample used in our analyses is composed of 77 sightlines, 75 of which have good column density ($\log N$) measurements or limits for \ion{Si}{4}, and all of them have good $\log N$ values or limits for \ion{O}{6} (Table \ref{sample}). 

\begin{table}
\begin{center}
\caption{Sample}
\label{sample}
\begin{tabular}{ccccccccc}
\hline
\hline
Ion & \multicolumn{3}{c}{${\mathcal{N}_{\rm star}}^a$} & \multicolumn{3}{c}{${\mathcal{N}_{\rm AGN}}^a$} & ${\sigma_p}^b$ & Ref. \\
\hline
\ion{Si}{4} & 49 & 13 & 13 & & $...$ & & 0.30 & \citetalias{Savage:2009aa} \\
 & & $...$ & & 119 & 11 & 0 & 0.13 & \citetalias{Zheng:2019aa} \\
\ion{O}{6} & 73 & 0 & 4 & & $...$ & & 0.23 & \citetalias{Savage:2009aa} \\
 & & $...$ & & 93 & 0 & 8 & 0.15 & \citetalias{Savage:2003aa} \\
\hline
\end{tabular}
\end{center}
{$^a$ These two columns are the number of sightlines from the stellar or AGN sample, respectively. In each column, three numbers are for sightlines with column density measurements, lower limits, and upper limits, respectively.}\\
{$^b$ $\sigma_p$ is the patchiness parameter (defined in Section 3.2), which represents the intrinsic scatter of the column density measurements in each sample. The patchiness parameter is derived to reduce the reduced $\chi^2$ value to $1$ for each sample individually.} \\
\vspace{-0.5cm}
\end{table}

For the AGN sample, we adopt two data sets.
We make use of the \ion{Si}{4} measurements from the COS-GAL sample \citepalias{Zheng:2019aa}, which is based on the Hubble Spectroscopic Legacy Archive \citep{Peeples:2017aa}.
Moreover, we retrieve the \ion{O}{6} measurements from the {\it FUSE} observations analyzed by \citetalias{Savage:2003aa}. 
We do not include the AGN sample in \citetalias{Savage:2009aa}, since it has a large overlap with the \citetalias{Zheng:2019aa} sample (18/25) and the \citetalias{Savage:2003aa} sample (22/25).
The final AGN sample includes 130 sightlines for \ion{Si}{4} and 101 sightlines for \ion{O}{6} (Table \ref{sample}).

\section{Models and Results}

\subsection{Previous Models}
Previously, the warm gas disk (e.g., traced by \ion{Si}{4} and \ion{O}{6}) of the MW is modeled as an 1-D plane-parallel slab model (\citealt{Jenkins:1978aa, Bowen:2008aa}; \citetalias{Savage:2009aa}).
The model only has one dimensional variation: the density distribution of the warm gas over the disk height $z$ as an exponential function of $n(z) = n_0 \exp(-|z|/z_0)$, where $n_0$ is the ion density at the mid-plane and $z_0$ is the scale height.
The current stellar sightlines are normally close to the Sun with a distance of $d \lesssim 2$ kpc, which implies that the average ion densities traced by these sightlines do not vary significantly at large scales.
Therefore, the stellar sightlines mainly determine the average density of the mid-plane around the solar system ($n_{\odot}$).
For the AGN sightlines, both the disk and the CGM are detected to show the large-scale variation.
Based on AGN sightlines, one could obtain the maximum projected column density along the $z$-direction ($N\sin |b|$) for the disk component, since in the plane-parallel slab model, the CGM contribution is ignored.
Combining these two measurements, the scale height $z_0$ in the plane-parallel slab model is derived as $N\sin |b|/n_{\odot}$ around the solar neighborhood.

This model works well for the sample dominated by stellar sightlines, such as \citetalias{Savage:2009aa}, which has $\approx 100$ stellar sightlines and $\approx 20$ AGN sightlines.
However, this model might have two problems with more and more AGN sightlines obtained by {\it HST}/COS.
First, for sightlines toward AGNs, the contribution from the MW CGM is not considered, which has been shown as an important component for low redshift galaxies ($z\lesssim 0.5$; e.g., \citealt{Stocke:2013aa, Werk:2014aa, Johnson:2015aa}).
Second, the AGN sightlines could trace the large scale variation of the disk in both the vertical and radial directions, so the plane-parallel slab model might lead to divergence from the observations.

The first problem is partially solved in \citetalias{Zheng:2019aa} by introducing an isotropic CGM component ($N_{\rm CGM}$) into the plane-parallel slab model; their model is referred as the two-component disk-CGM model hereafter. 
They applied the two-component disk-CGM model to fit the \ion{Si}{4} column density distribution measured along 130 AGN sightlines across the Galactic sky, and found a significant contribution of the MW CGM of $\log N_{\rm CGM} \approx 13.53$. 
The \citetalias{Zheng:2019aa} analyses provide the first statistical evidence that the MW hosts an extended warm CGM.
However, in this model, the disk component ($\log N_{\rm disk} = 12.1$) is different from \citetalias{Savage:2009aa} ($\log N_{\rm disk} = 13.4$) by more than one order of magnitude.
Therefore, there is still a huge gap between the model dominated by stellar sightlines (the plane-parallel slab model; \citetalias{Savage:2009aa}) and the model dominated by AGN sightlines (the two-component disk-CGM model; \citetalias{Zheng:2019aa}).
In the following, we introduce a 2-D disk-CGM model with a disk radial profile that alleviates the tension between the flat-slab model by \citetalias{Savage:2009aa} and the two-component disk-CGM model by \citetalias{Zheng:2019aa} in studying the warm gas in the MW.

\begin{table*}
\begin{center}
\caption{The Disk and CGM model for \ion{Si}{4} and \ion{O}{6}}
\label{fits}
\begin{tabular}{lcccccccccc}
\hline\hline
Model & $\log n_0$ & $r_0$ & $z_0$ & $\log N_{\rm mp}^{\rm CGM}$ & $\log N_{\rm nd}^{\rm CGM}$ & red. $\chi^2$ (dof) & $\log n_{\odot}^{\rm disk}$ & $\log n_\odot z_0$ & $\log M_{\rm disk}$\\
 & $(\rm cm^{-3})$ & kpc & kpc & $(\cmsq)$ & $(\cmsq)$ & & $(\cc)$ & $(\cmsq)$ & $(M_\odot)$\\
 (1) & (2) & (3) & (4) & (5) & (6) & (7) & (8) & (9) & (10)\\

\hline
\multicolumn{10}{c}{The 2-D disk-CGM models with isotropic CGM of \ion{Si}{4}}\\
\hline
$\rm R_EZ_E$ & $-7.82\pm0.24$ & $4.3\pm 1.2$ &  $2.5\pm0.6$ & \multicolumn{2}{c}{$13.18\pm0.12$} & 1.232 (201) & -8.68 & 13.21 & 3.78 \\
$\rm R_GZ_E$ & $-8.21\pm0.16$ & $8.3\pm 1.5$ &  $2.8\pm0.7$ & \multicolumn{2}{c}{$13.14 \pm 0.15$} & 1.182 (201) & -8.67 & 13.26 & 3.70 \\
$\rm R_EZ_G$ & $-7.74\pm0.23$ & $3.7\pm 0.9$ &  $3.3\pm0.6$ & \multicolumn{2}{c}{$13.16 \pm 0.11$} & 1.267 (201) & -8.73 & 13.22 & 3.80 \\
$\rm R_GZ_G$ & $-8.21\pm0.15$ & $7.4\pm 1.1$ &  $3.5\pm0.7$ & \multicolumn{2}{c}{$13.12 \pm 0.14$} & 1.202 (201) & -8.78 & 13.26 & 3.66 \\
\hline
\multicolumn{10}{c}{The 2-D disk-CGM models with isotropic CGM of \ion{O}{6}}\\
\hline
$\rm R_EZ_E$ & $-7.19\pm0.20$ & $5.7\pm1.8$ &  $2.3\pm0.6$ & \multicolumn{2}{c}{$13.98\pm0.14$} & 1.210 (174) & -7.84 & 14.02 & 4.39 \\
$\rm R_GZ_E$ & $-7.51\pm0.13$ & $9.8\pm 1.9$ &  $2.6\pm0.7$ & \multicolumn{2}{c}{$13.93 \pm 0.18$} & 1.196 (174) & -7.83 & 14.08 & 4.29 \\
$\rm R_EZ_G$ & $-7.18\pm0.20$ & $5.4\pm 1.6$ &  $2.4\pm0.5$ & \multicolumn{2}{c}{$14.03 \pm 0.09$} & 1.240  (174) & -7.86 & 13.95 & 4.31 \\
$\rm R_GZ_G$ & $-7.50\pm0.12$ & $8.8\pm 1.4$ &  $2.7\pm0.6$ & \multicolumn{2}{c}{$13.98 \pm 0.12$} & 1.222 (174) & -7.90 & 14.01 & 4.16 \\
\hline
\multicolumn{10}{c}{The 2-D disk-CGM models with anisotropic CGM of \ion{Si}{4}}\\
\hline
$\rm R_EZ_E$ & $-7.93\pm0.20$ & $5.2\pm 1.4$ &  $2.6\pm0.6$ & $12.62\pm0.40$ & $13.32\pm0.08$ & 1.128 (200) & -8.64 & 13.27 & 3.86 \\
$\rm R_GZ_E$ & $-8.25\pm0.14$ & $9.2\pm 1.6$ &  $2.7\pm0.7$ & $12.67 \pm 0.44$ & $13.28\pm0.10$ & 1.106 (200) & -8.62 & 13.29 & 3.74 \\
$\rm R_EZ_G$ & $-7.89\pm0.19$ & $4.6\pm 1.0$ &  $3.4\pm0.6$ & $12.46 \pm 0.45$ & $13.32 \pm 0.07$ & 1.138  (200) & -8.69 & 13.28 & 3.86 \\
$\rm R_GZ_G$ & $-8.27\pm0.12$ & $8.4\pm 1.2$ &  $3.5\pm0.7$ & $12.48 \pm 0.52$ & $13.28 \pm 0.08$ & 1.105 (200) & -8.72 & 13.31 & 3.70 \\
\hline
\multicolumn{10}{c}{The 2-D disk-CGM models with anisotropic CGM of \ion{O}{6}}\\
\hline
$\rm R_EZ_E$ & $-7.35\pm0.15$ & $8.0\pm2.5$ &  $2.6\pm0.6$ & $13.21\pm0.58$ & $14.19\pm0.08$ & 1.104 (173) & -7.82 & 14.08 & 4.55 \\
$\rm R_GZ_E$ & $-7.60\pm0.09$ & $12.2\pm 2.4$ &  $2.7\pm0.6$ & $13.18 \pm 0.66$ & $14.16\pm0.10$ & 1.108 (173) & -7.81 & 14.11 & 4.40 \\
$\rm R_EZ_G$ & $-7.33\pm0.15$ & $7.2\pm 2.2$ &  $2.9\pm0.5$ & $13.21 \pm 0.50$ & $14.22 \pm 0.07$ & 1.121 (173) & -7.84 & 14.05 & 4.49 \\
$\rm R_GZ_G$ & $-7.59\pm0.09$ & $10.7\pm 1.8$ &  $3.0\pm0.6$ & $13.18 \pm 0.57$ & $14.19 \pm 0.08$ & 1.119 (173) & -7.84 & 14.08 & 4.28 \\
\hline
\end{tabular}
\end{center}
Notes: Column 1) The model name: R and Z denote radial and vertical directions; E and G denote exponential and Gaussian profiles. Column 2) The density at the GC. Column 3) The scale length. Column 4) The scale height. Column 5) The CGM column density along the disk mid-plane (`mp'). For isotropic CGM model, this value is the same as the CGM column density perpendicular to the disk. Column 6) The CGM column density perpendicular to the disk (`nd' denotes the normal line). Column 7) The reduced $\chi^2$ and the degree of freedom. Column 8) The ion density at the solar system. Column 9) The disk column density toward $b=90^\circ$ at the solar system. Column 10) The disk mass of ions.
\end{table*}

\subsection{The 2-D Disk-CGM Model}

We improve the previous models by introducing a two-dimension (2-D) disk into the two-component disk-CGM model of \citetalias{Zheng:2019aa}.
In this model, we consider the number density distribution of the disk component, which is a 2-D distribution ($n_{\rm disk}(r, z)$) depending on the radius ($r$) from the Galactic center (GC), and the $z$-height above and below the Galactic plane.
For a given sightline at a given distance ($l$, $b$, and $d$), we can calculate the column density contribution from the disk by integrating the 2-D density distribution of the disk component. 
For the CGM component, we first consider a constant CGM column density over all directions (isotropic $N_{\rm CGM}$; the same as \citetalias{Zheng:2019aa}).
This CGM component is only applied to the AGN sightlines, while the disk component is calculated for both stellar and AGN sightlines.
For a given sightline, the model predicted column densities are
\begin{eqnarray}
N(l, b, d) &=& N_{\rm disk}(l, b, d) \rm~ for~ stars, \notag \\
N(l, b) &=& N_{\rm disk}(l, b, d_{\rm max}) + N_{\rm CGM} ~\rm for~ AGNs,
\end{eqnarray}
where $d_{\rm max}$ is the maximum distance for the disk component, which is set to be the virial radius of the MW halo ($R_{\rm vir} = 250$ kpc).

{Here, we emphasize that the decomposition of the disk and the CGM component is phenomenological, since we assume the stellar sightlines do not trace any CGM gas.
This is limited by the current sample, which does not have sightlines in the MW halo that trace the radial profile of the MW CGM at large radii, so we cannot calculate the CGM contribution to the column density measurements in stellar sightlines.}
However, this assumption is also reasonable with the current sample.
For stellar sightlines, most stars are close to the disk mid-plane center $|z| \lesssim 3~\rm kpc$, which are marginally affected by the CGM component.
There are only three stars have $|z| > 3~\rm kpc$, leading a tiny effect on the fitting results.

For the 2-D disk component, the radial and vertical profiles are assumed to be independent from each other, so the ion number density distribution in the disk is
\begin{equation}
\label{disk_model}
n(r, z) = n_0 f_r(r) f_z(z),
\end{equation}
where $f_r(r)$ and $f_z(z)$ are the profile functions in the radial and vertical directions.
For $f_z(z)$, we adopt the same exponential profile, $f_z(z)=\exp(-|z|/z_0)$, as the plane-parallel slab model \citepalias{Savage:2009aa}. 
We also assume the radial profile to be exponential as $f_r(r)=\exp(-r/r_0)$, where $r_0$ is the scale length.
Both of the radial and vertical exponential profiles are empirical as inferred from the \ion{H}{1} disk and the stellar disk \citep{Kalberla:2008aa, Bovy:2013aa}.
It is possible that the warm gas disk follows a different density distribution, since the warm gas disk is more extensive and affected by Galactic feedback.
Therefore, in our following analyses, we also consider the Gaussian function $f(x) = \exp(-(x/x_0)^2)$ for both $f_r(r)$ and $f_z(z)$ to test whether the shape of the warm gas disk can be distinguished from the observations.
In total, there are four phenomenological models for our disk density profiles, named as $\rm R_E Z_E$ (exponential radial and vertical profiles), $\rm R_G Z_E$ (Gaussian radial profile and exponential vertical profile), $\rm R_E Z_G$ (Exponential radial profile and Gaussian vertical profile), and $\rm R_G Z_G$ (Gaussian radial and vertical profiles). 
In these models, the solar system is placed at $r_\odot=8.5\rm~kpc$ \citep{Ghez:2008aa} and $z_\odot=0\rm~kpc$.

We apply these models to the column density measurements of \ion{Si}{4} and \ion{O}{6} \citepalias{Savage:2003aa, Savage:2009aa, Zheng:2019aa}, and obtain the best parameters using the minimum $\chi^2$ estimation.
In our fittings, we include the lower or upper limits of $\log N_{\rm SiIV}$ and $\log N_{\rm OVI}$ values, which are typically not considered in previous modelings (e.g., \citetalias{Savage:2009aa}).
For these limits, we only calculated the $\chi^2$ value when it is opposite to the limits, i.e., higher than the upper limit and lower than the lower limit; otherwise, the $\chi^2$ value is fixed to 1 for these limits.
The uncertainties of the sightlines are one-sided uncertainty, which we set according to the intrinsic scatters (i.e., the patchiness parameter derived later; Table \ref{sample}).
Therefore, the uncertainty of limits in the stellar sample is set as 0.3 dex, while the AGN sample is set as 0.1 dex.

Previous studies show that the intrinsic column density scatters of the disk and the CGM are the major contributors of the deviation in the fitting (i.e., \citealt{Bowen:2008aa}; \citetalias{Savage:2009aa, Zheng:2019aa}).
The intrinsic scatter is modeled as the patchiness parameter ($\sigma_{\rm p}$), which is an additional uncertainty attached to the measurement uncertainty as $\sigma^2_{\rm f} = \sigma^2_{\rm p} + \sigma^2_{\rm m}$.
$\sigma_{\rm f}$ is the final adopted uncertainty in model fittings, and $\sigma_{\rm m}$ is the measurement uncertainty.
There are two methods to implement patchiness parameters.
The first method is varying the patchiness parameter to obtain the reduced $\chi^2 = 1$, which is adopted in \citetalias{Savage:2009aa} (and reference herein).
The second method is to implement the patchiness uncertainty into the Bayesian model as introduced in \citetalias{Zheng:2019aa}.
These two methods obtain similar results, and we adopt the first method in our analyses.
The calculated $\sigma_{\rm p}$ values are shown in Table \ref{sample}, which leads to reduced $\chi^2$ of $0.95-1.05$ due to the significant figures.
In \citetalias{Savage:2009aa}, the patchiness parameters of \ion{Si}{4} and \ion{O}{6} are 0.266 and 0.233, respectively.
The larger patchiness parameters (0.30 and 0.23) in Table \ref{sample} are mainly due to the inclusion of upper or lower limits in our fittings, and the exclusion of AGN sightlines, which normally have smaller scatters.
For the AGN samples, the previously $\sigma_{\rm p}$ are 0.18 and 0.25 for \ion{Si}{4} and \ion{O}{6}, respectively \citepalias{Savage:2003aa, Zheng:2019aa}, which are larger than our values in Table \ref{sample} (0.13 and 0.15).
The reduction of the patchiness parameter indicates that the radial distribution of the disk affects the AGN sample more significantly.

\begin{figure*}
\begin{center}
\subfigure{
\includegraphics[width=0.32\textwidth]{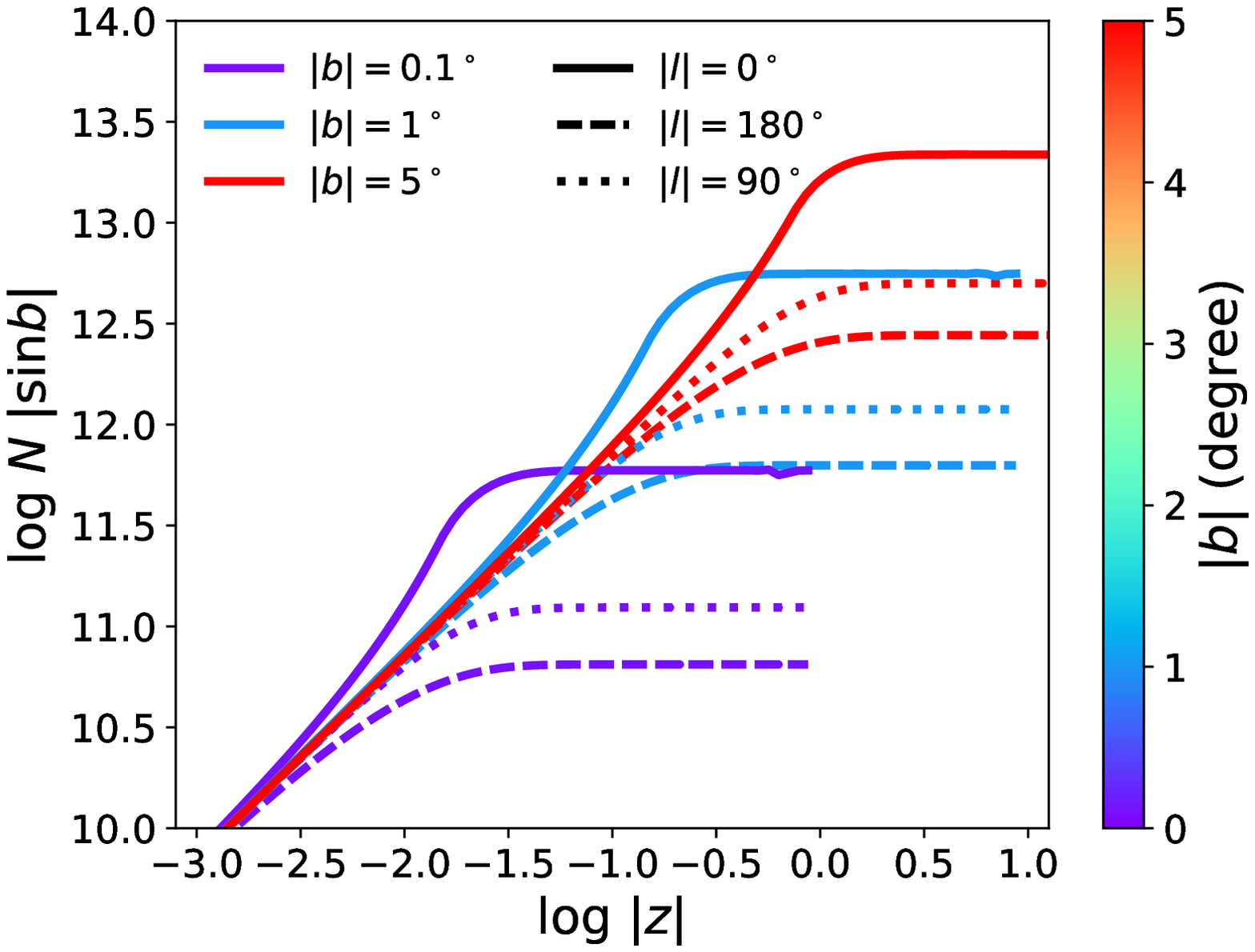}
\includegraphics[width=0.32\textwidth]{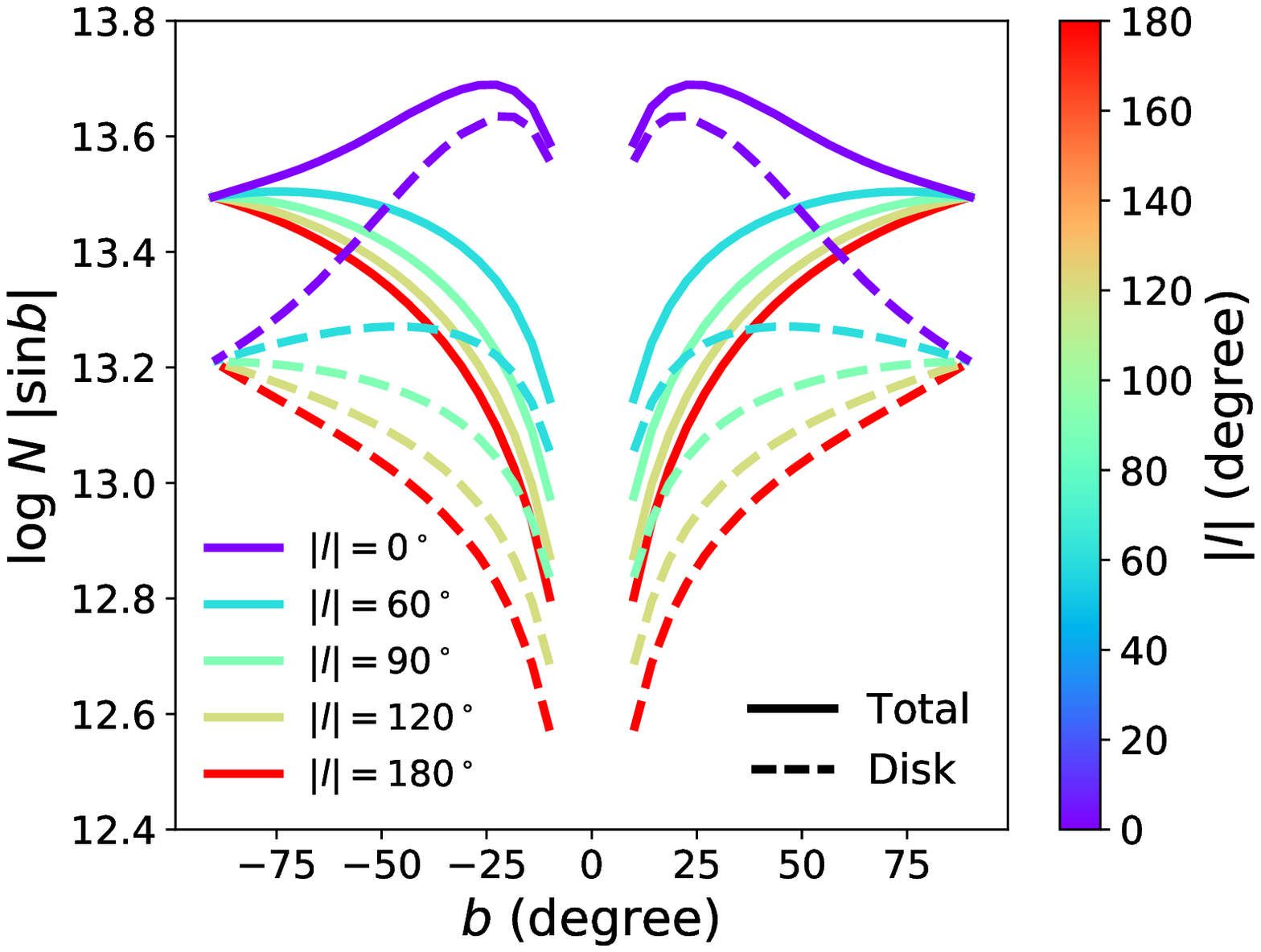}
\includegraphics[width=0.32\textwidth]{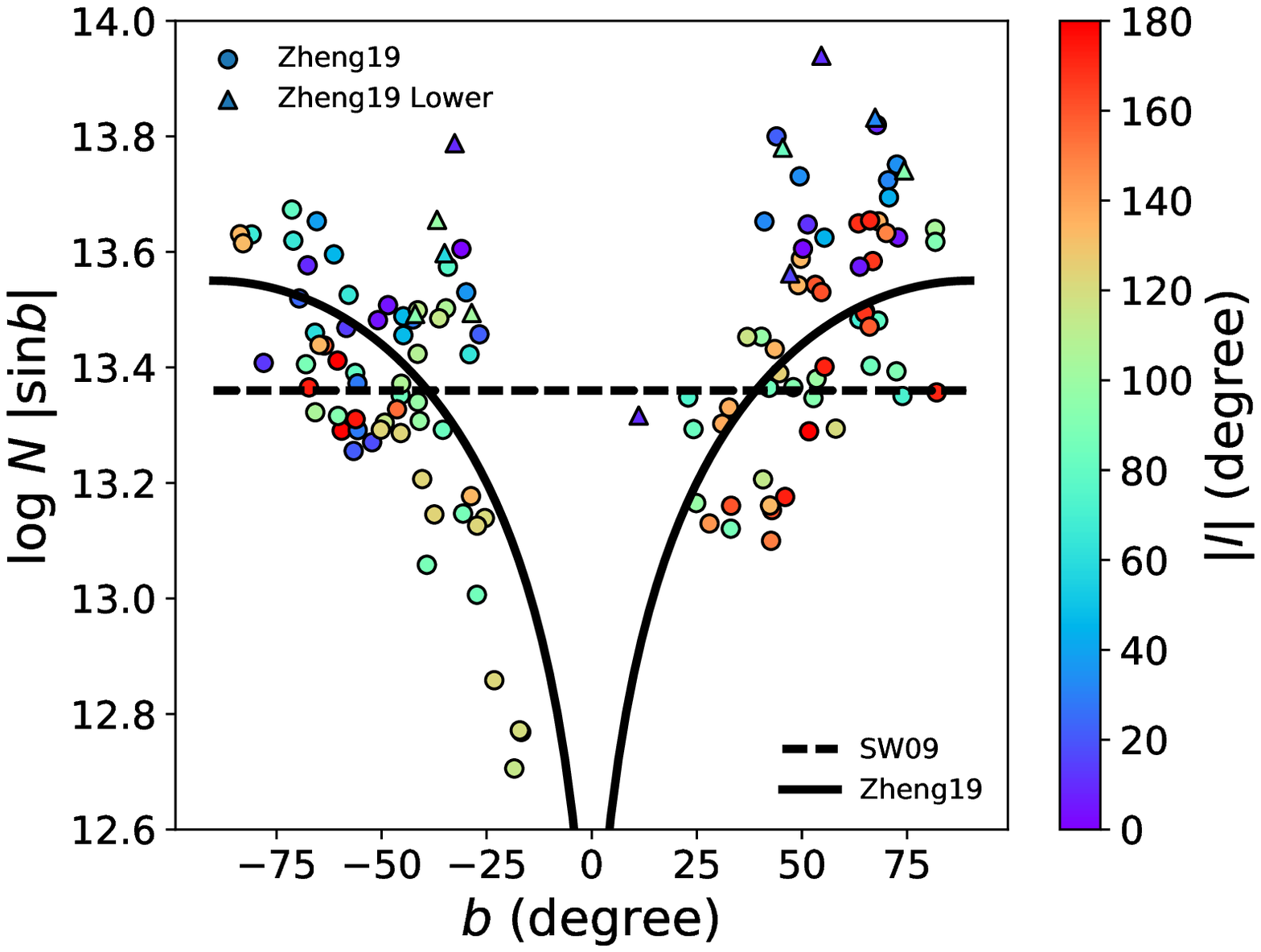}
}
\caption{The prediction $\log N \sin|b|$ using $\rm R_EZ_E$ model with the isotropic CGM. The model parameters are adopted for \ion{Si}{4} (first row of Table \ref{fits}). {\it Left panel}: the predicted relationship between $\log N \sin|b|$ and $\log |z|$ for stellar sightlines at $|b|< 5^\circ$ with contribution only from the disk component.  {Lower latitudes and larger longitudes lead to lower projected column densities ($N\sin |b|$), hence lower observed scale heights ($N\sin |b|/n_\odot$; see the definition in the text), which is proportional to the projected column density}. {\it Middle panel}: the predicted relationship between $\log N \sin|b|$ and $\log |b|$ for AGN sightlines at $|b| \gtrsim 30^\circ$. The dashed lines are the disk component, while the solid lines are the total model. Sightlines around the anti-GC show rapid decrease of the projected column densities with lower latitudes due to the disk radial variation. {\it Right panel}: the plane-parallel slab model (\citetalias{Savage:2009aa}; the dashed line) and the two-component disk-CGM model  (\citetalias{Zheng:2019aa}; the solid line). In these two models, $\log N \sin|b|$ only has dependence on Galactic latitude, since these two models only have 1-D disk with a density profile over the height-$|z|$.}
\label{model}
\end{center}
\end{figure*}

The fitting results are summarized in Table \ref{fits}.
Overall, the exponential function leads to smaller scale lengths or scale heights than the Gaussian function, because the exponential function has a slower decay with the same characteristic length.
With the isotropic CGM, the $\rm R_GZ_E$ model (indicating Gaussian function for the radial profile and exponential function for the vertical profile) is preferred with the significance of $\lesssim 2 \sigma$ (inferred from the difference of total $\chi^2 \lesssim 4$ ).
Similarly, the anisotropic CGM models show that no specific model is preferred, which will be described and discussed in detail in Section 3.3.
Therefore, we suggest that the current stellar and AGN samples cannot distinguish the density profiles (exponential or Gaussian) of the warm gas disk, and we set the $\rm R_EZ_E$ model as the fiducial model.

All of our four models show that both the disk and the CGM components contribute significantly to the observed column densities in AGN sightlines.
Using $n_0z_0$ as the characteristic column density of the disk, the disk component is comparable to the CGM component for both \ion{Si}{4} and \ion{O}{6} (Table \ref{fits}).
Previously, the plane-parallel slab model shows that the disk component has \ion{Si}{4} column density ranging from $\log N = \log n_0z_0$ = 13.36 to 13.56, and \ion{O}{6} from $\log N$ =14.12 to 14.28 \citepalias{Savage:2009aa}.
These values are all larger than our values of $13.21-13.26$ (\ion{Si}{4}) and $13.95-14.08$ (\ion{O}{6}).
The lower values of our disk component are because we take into account the contribution of the MW CGM to ion column density measurements toward AGN sightlines, whereas the plane-parallel slab model assumes no contribution from the CGM.
Our fitting results show that the CGM components are $13.12-13.17$ (\ion{Si}{4}) and $13.91-14.02$ (\ion{O}{6}).
These values are comparable with the \ion{Si}{4} and \ion{O}{6} column densities measured from transverse AGN sightlines at $R\approx 100\rm~kpc$ ($\approx 0.5 R_{\rm vir}$) for low redshift $L^*$ galaxies ($z\approx 0.5$; \citealt{Werk:2013aa, Savage:2014aa, Johnson:2015aa}).
{Although the sightlines through the MW CGM have a different geometry from sightlines for external galaxies, this consistency indicates a decreasing column density dependence on the radius of a power law with a slope of about $-1$ \citep{Werk:2013aa}.}

For the disk component, although we cannot distinguish between the exponential and Gaussian profiles, the scale height ($z_0$) and the scale length ($r_0$) can be determined.
In the fiducial model ($\rm R_EZ_E$), the scale heights are $2.6\pm0.6$ kpc and $2.4\pm 0.6$ kpc for \ion{Si}{4} and \ion{O}{6}, respectively.
The scale lengths are $4.2\pm1.2$ kpc and $5.6\pm 1.7$ for \ion{Si}{4} and \ion{O}{6}, respectively.
The scale lengths are first measured in this work for the MW warm gas disk.

The radial profile of the disk component is important to solve the divergence between the plane-parallel slab model \citepalias{Savage:2009aa} and the two-component disk-CGM model \citepalias{Zheng:2019aa}, which have different relative contributions between disk and CGM components.
In Fig. \ref{model}, we predict the projected column density distribution ($\log N \sin|b|$) as a function of $|z|$-height and Galactic latitude ($b$) using the $\rm R_EZ_E$ model with isotropic CGM with best-fit parameters in Table \ref{fits}.

Before introducing the plots, we define the observed scale height, which is an observable for the warm gas analysis.
This parameter is defined as $N\sin |b|/n_\odot$, where $N \sin |b|$ is the projected column density observed from the Sun, and $n_\odot$ is the mid-plane ion density around the Sun.
The observed scale height could be estimated as the $z$-height of the turnover point in the projected column density function of $z$-height.
For example, in the left panel of Fig. \ref{model}, the observed scale height is about $|z_{\rm obs}| \approx 0.03$ kpc at $|b|=0.1^\circ$, while it is $|z_{\rm obs}| \approx 1$ kpc at $|b|=5^\circ$.
This is different from the scale height $z_0$ defined in Equation (\ref{disk_model}), which is a constant over the entire sky.
The scale height $z_0$ could be calculated as $N_{\rm r,nd}/n_{\rm r, mp}$, where $N_{\rm r,nd}$ and $n_{\rm r, mp}$ are the column density toward $b=90^\circ$ (the normal direction of the disk) and the mid-plane density at any given radius of $r$.

The difference between the scale height ($z_0$) and the observed scale height ($z_{\rm obs}$) is mainly due to the radial density distribution of the disk component.
Using two AGN sightlines as an example, one sightline is toward $b=90^\circ$, while another sightline is toward the anti-Galactic center (anti-GC; $l=180^\circ$) direction with any Galactic latitudes.
Considering the calculations of the scale height and the observed scale height, the mid-plane densities are the same, since both of densities are around the Sun.
However, the projected column densities of the disk component are different: the term $N_{r_\odot, {\rm nd}}$ is always larger than $N \sin |b|$ at different $b$.
This is a result of the disk radial distribution, since $N \sin |b|$ could be approximated as $N_{r_\odot, {\rm nd}} \exp(-(r-r_\odot)/r_0)$, where $r$ is always larger than $r_\odot$ for anti-GC sightlines.
Therefore, we expected that the observed scale height is always lower than the real scale height for anti-GC sightlines.
Also, since a low $b$ leads to a small $\tan b$ value, the low latitude sightlines need a longer path length to reach the same height.
Then, the effect of the disk radial distribution is more significant for low latitude sightlines, and lower projected column densities are expected for these sightlines (the left panel in Fig. \ref{model}).

The sightlines toward the GC ($l=0^\circ$) is more complex, since the disk radial distribution leads to higher density around the GC.
However, the stellar sightlines are mainly at low latitudes $|b|\lesssim 5^\circ$.
In our fiducial $\rm R_E Z_E$ model, the scale heights of both \ion{Si}{4} and \ion{O}{6} are higher than 2 kpc.
Using this scale height and a Galactic latitude of $5^\circ$, one expects a radius difference of $z_0/ \tan |b| >20$ kpc to reach the scale height of the disk.
With this radius difference, the final effect on the observed scale height will be a competition between the high-density gas around the GC and the low-density gas at large radii.
Our numerical calculation shows that it is possible to have larger observed scale heights around the GC direction (the left panel in Fig. \ref{model}).
Therefore, the sightlines around the GC direction have higher projected column densities than those toward anti-GC directions due to the high-density gas at the GC.
Since the solar system is at $r_\odot=8.5$ kpc, this difference is most significant around $|z|=8.5 \tan |b| ~\rm kpc$, which is $\approx 0.7$ kpc for $|b|=5^\circ$.

In the middle panel of Fig. \ref{model}, we show the predicted projected column density for the AGN samples using the fiducial $\rm R_E Z_E$ model.
Our model predicts that the projected column density has a dependence on both Galactic latitude and Galactic longitude, while the previous models only have dependence on Galactic latitude (\citetalias{Savage:2009aa} and \citetalias{Zheng:2019aa}; the right panel of Fig. \ref{model}).
The $\log N$ dependence on both $l$ and $b$ is due to the radial profile of the disk component, so it is similar to the case in the disk-only model (the left panel of Fig. \ref{model}), but for higher Galactic latitudes ($|b|>30^\circ$).
The projected column densities are generally higher toward the GC direction than the anti-GC direction, and all the values converge at $|b|=90^\circ$.
The anti-GC sightlines show a more significant dependence on the Galactic latitude $|b|$ (decreasing rapidly), which is due to the radial profile in our model.
The \citetalias{Zheng:2019aa} model also reproduces this feature, but in a different way.
The Galactic latitude dependence in the \citetalias{Zheng:2019aa} model is due to the term of $N_{\rm CGM}\sin |b|$, and do not have dependence on the Galactic longitude, so this model does not reproduce the feature that the GC sightlines have higher column density than the anti-GC sightlines (also see Fig. 5 in \citetalias{Zheng:2019aa}, and Fig. 7 in \citealt{Wakker:2012aa}).

\subsection{The Anisotropy of the CGM Component}
In the previous section, we adopt the isotropic CGM assumption from \citetalias{Zheng:2019aa}.
However, this isotropic CGM profile over the entire sky may not best represent the gas density distribution in the CGM.
For example, \citet{Bordoloi:2011aa} found that the absorption features are stronger along the minor axis using \ion{Mg}{2} absorption lines for external galaxies (also see \citealt{Lan:2018aa}).
Besides the absorption strength, \citet{Martin:2019aa} found that the non-detections of \ion{Mg}{2} are mainly along the minor axis (perpendicular to the disk), which indicates a lower detection rate along the minor axis.
Therefore, we consider the azimuthal variation of CGM in our 2-D disk-CGM model of the MW.

\begin{figure*}[!t]
\begin{center}
\subfigure{
\includegraphics[width=0.48\textwidth]{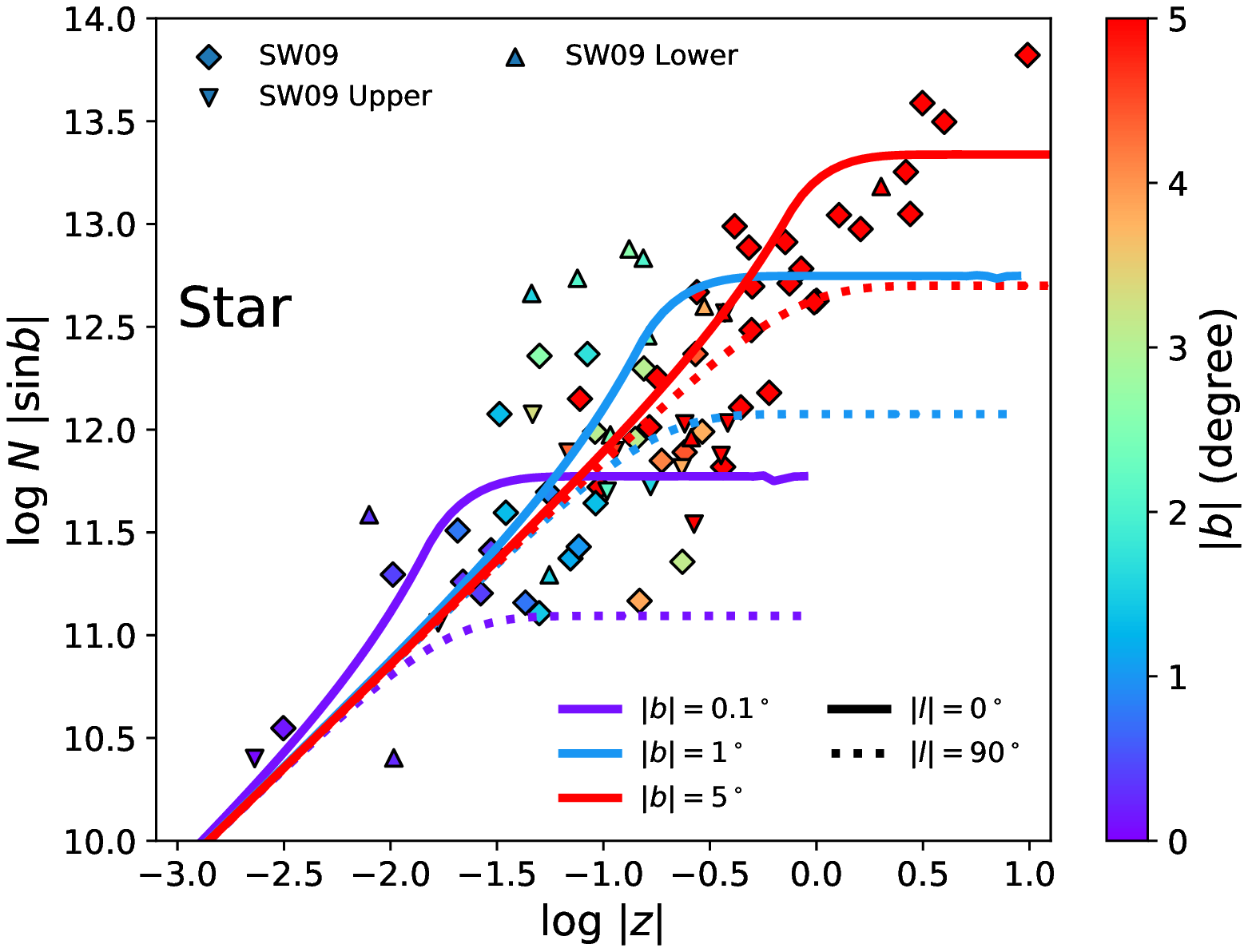}
\includegraphics[width=0.48\textwidth]{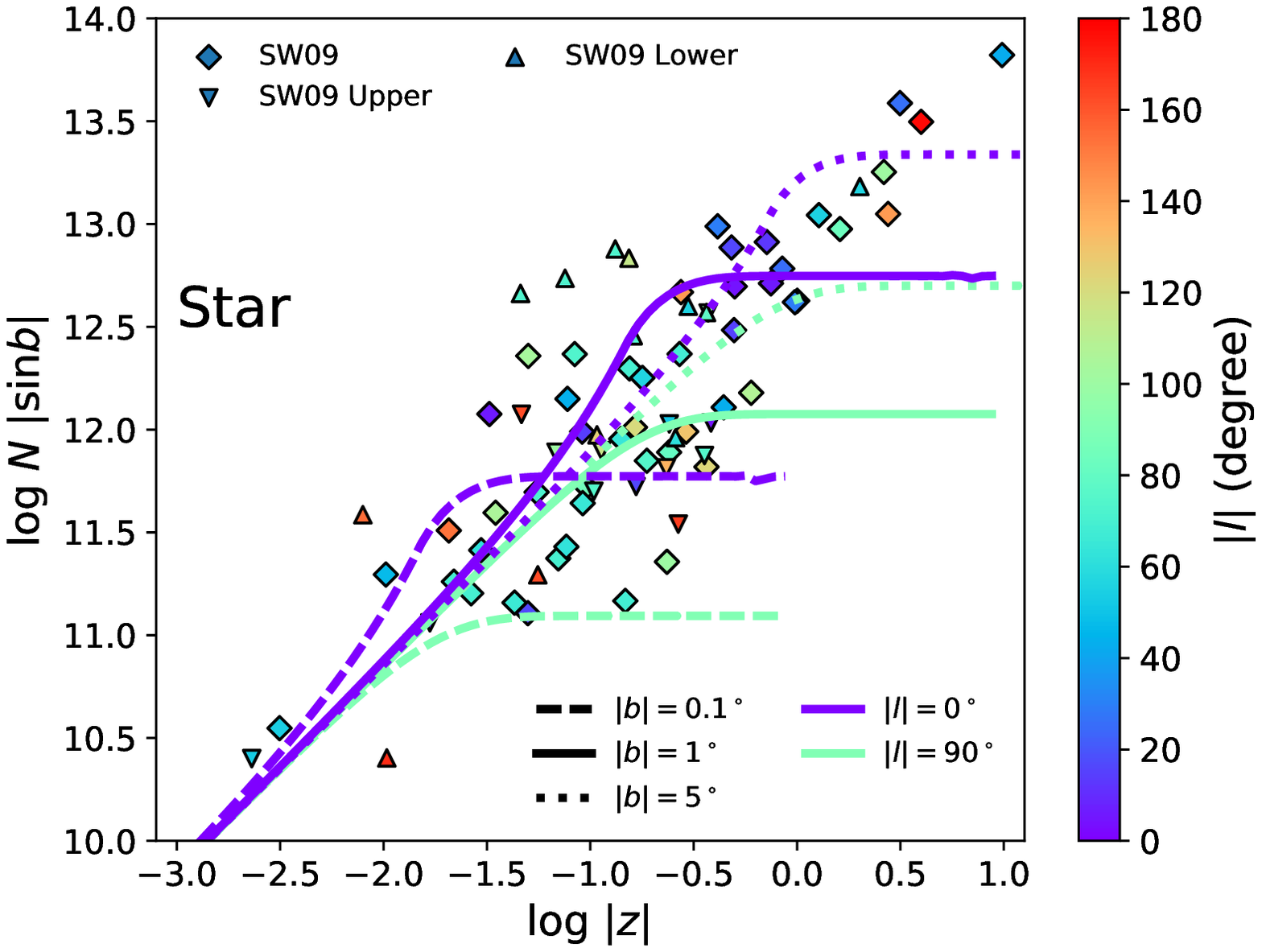}
}
\subfigure{
\includegraphics[width=0.48\textwidth]{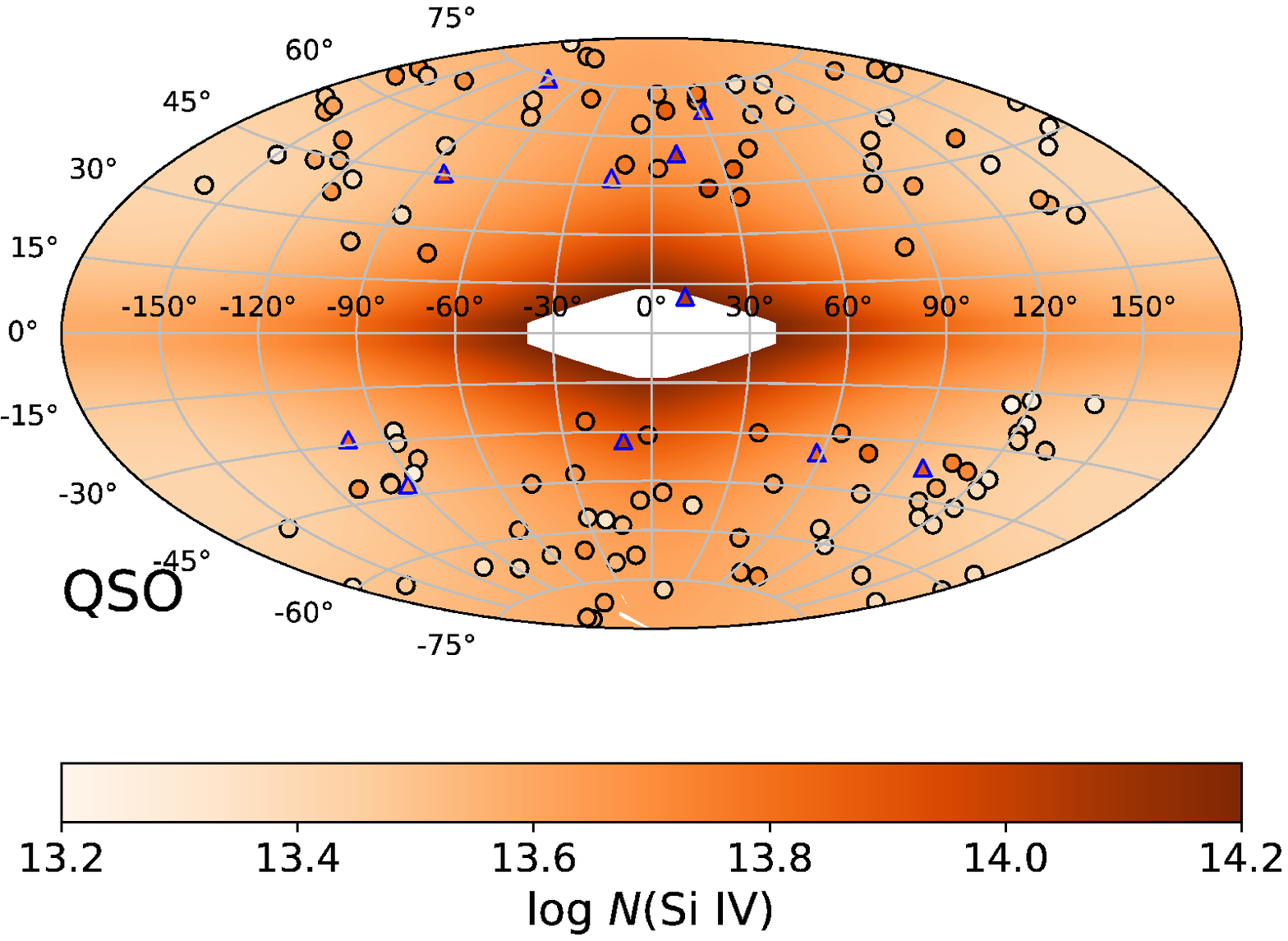}
\includegraphics[width=0.48\textwidth]{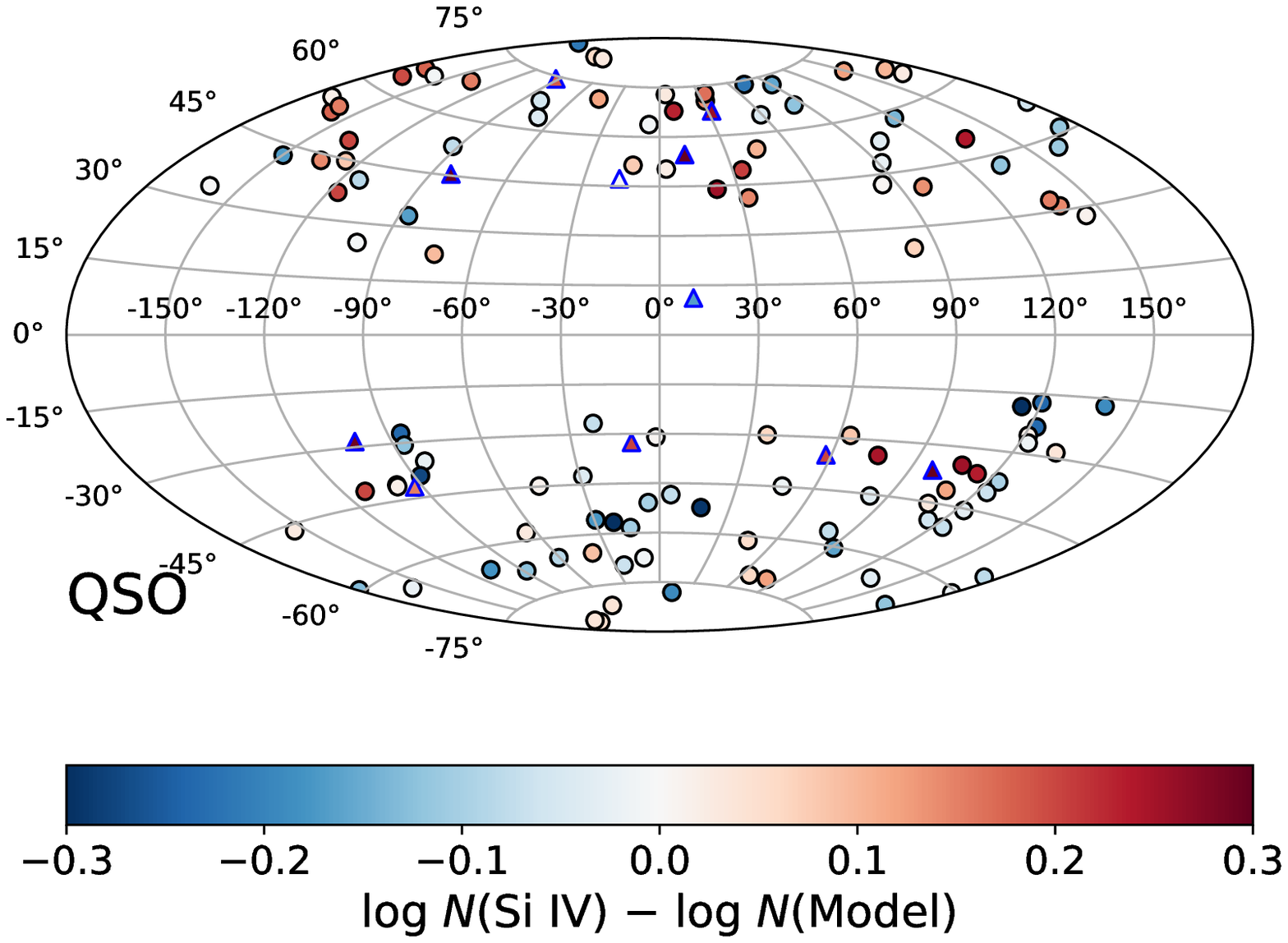}
}
\subfigure{
\includegraphics[width=0.48\textwidth]{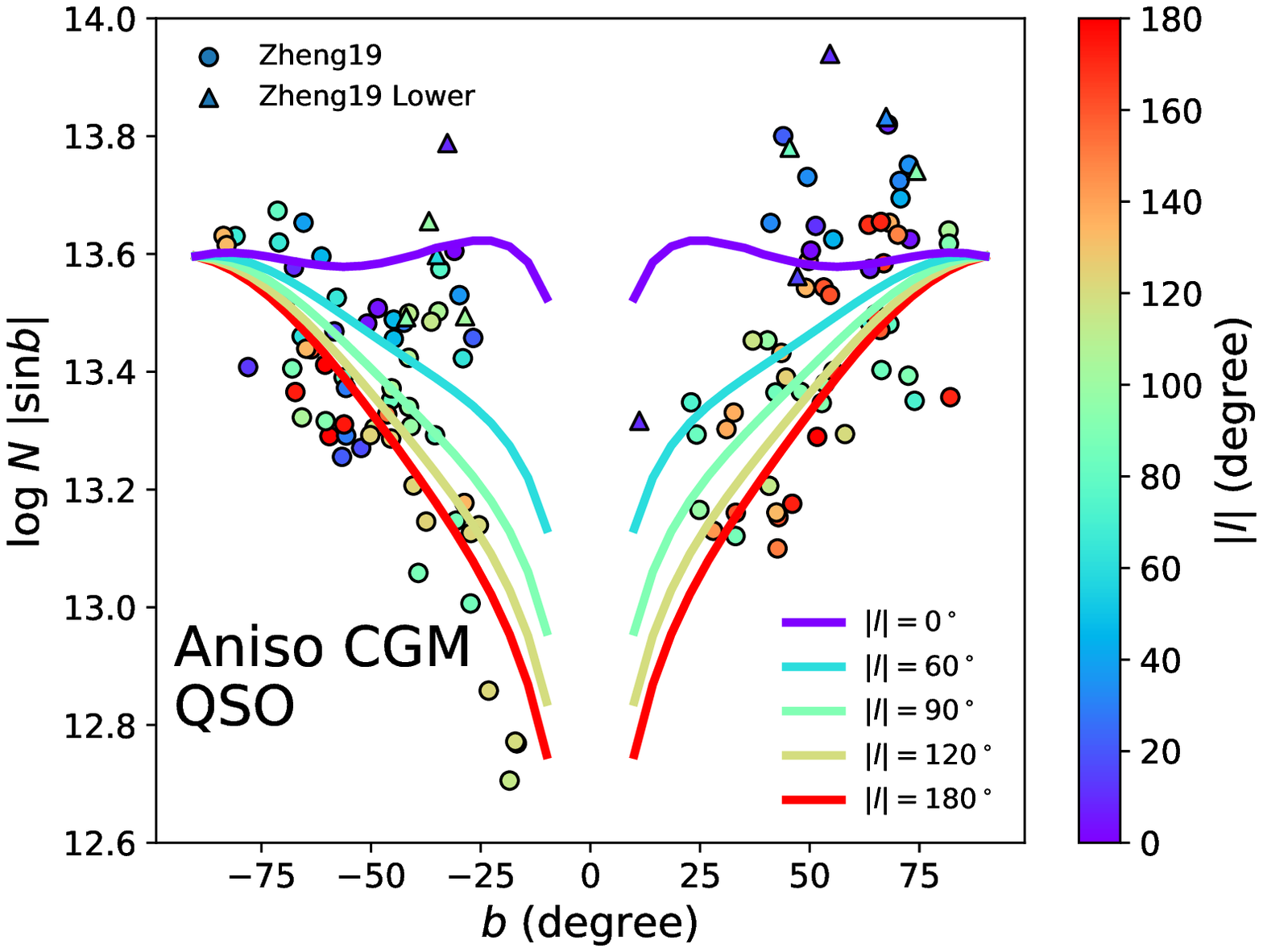}
\includegraphics[width=0.48\textwidth]{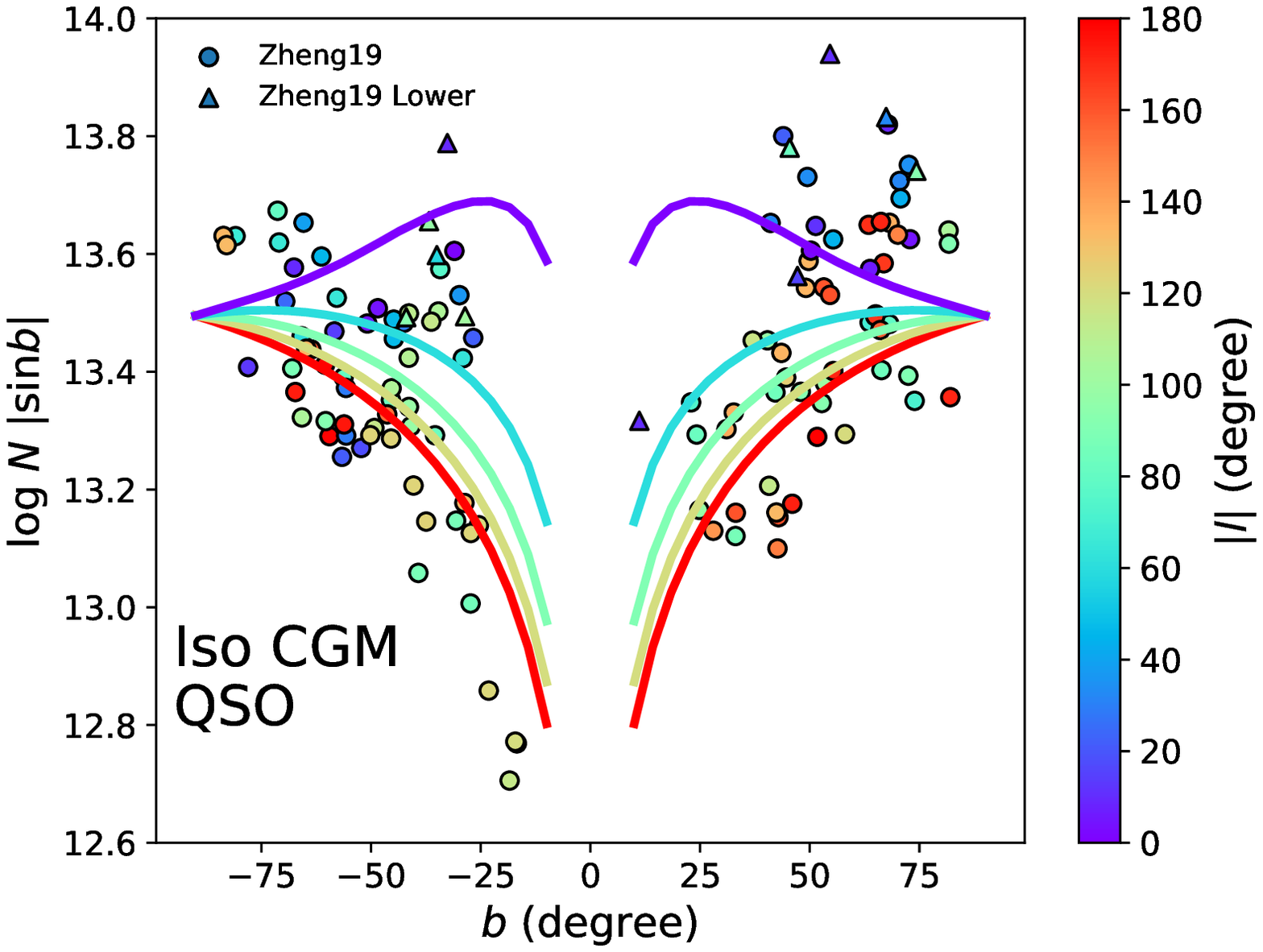}
}
\caption{The comparison between 2-D disk-CGM model predictions and observations for \ion{Si}{4}. {\it Top panels}: the comparison for the stellar sample. Two plots are color-coded in Galactic latitude ($|b|$; left) and Galactic longitude ($|l|$; right), respectively. Lower $|b|$ sightlines have lower projected column densities since these sightlines are more affected by the disk radial distribution (need longer path length to reach the same height). Sightlines toward the GC have higher projected column densities due to the high ion density around the GC. {\it Middle panels}: the global variation of total column densities for the AGN sample plotted in the Aitoff projection (the left panel). {In the left panel, the white diamond-like region at the GC ($l=0^\circ$ and $b=0^\circ$) has column densities of $\log N>14.2$, so it is left as a blank region. The model predicts that the minimum column density for AGN sightlines occurs around Galactic latitudes of $30^\circ-50^\circ$,} which is a result of the competition between the disk component (the minimum at $b=90^\circ$) and the CGM component (the minimum at $b=0^\circ$). The right panel is the residual of $\log N(\rm Observation) - \log N(\rm Model)$, {which mainly shows the north-south asymmetry (discussed in Section 4.4)}. {\it Lower panels}: the comparison between anisotropic CGM model (left panel) and the isotropic CGM model (right panel). The anisotropic CGM model is $4.6\sigma$ better than the isotropic CGM model by reducing the total $\chi^2$ of 20.8. The anisotropic CGM model reproduces the sharp decreasing of the projected column density at low Galactic latitudes better for sightlines toward the anti-GC ($|l|=180^\circ$; also see Fig. \ref{OVI_plots}).}
\label{SiIV_plots}
\end{center}
\end{figure*}

\begin{figure*}[!t]
\begin{center}
\subfigure{
\includegraphics[width=0.48\textwidth]{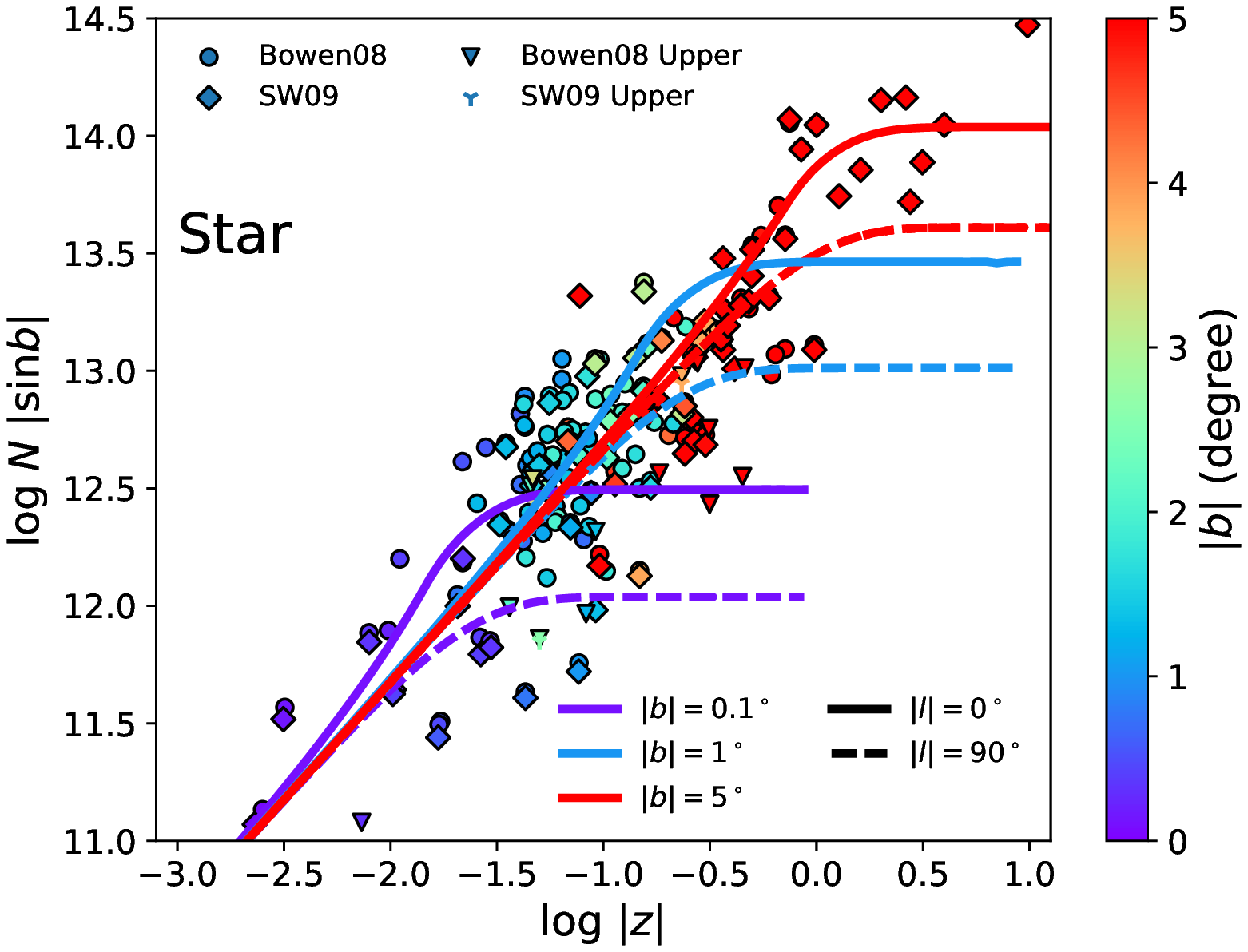}
\includegraphics[width=0.48\textwidth]{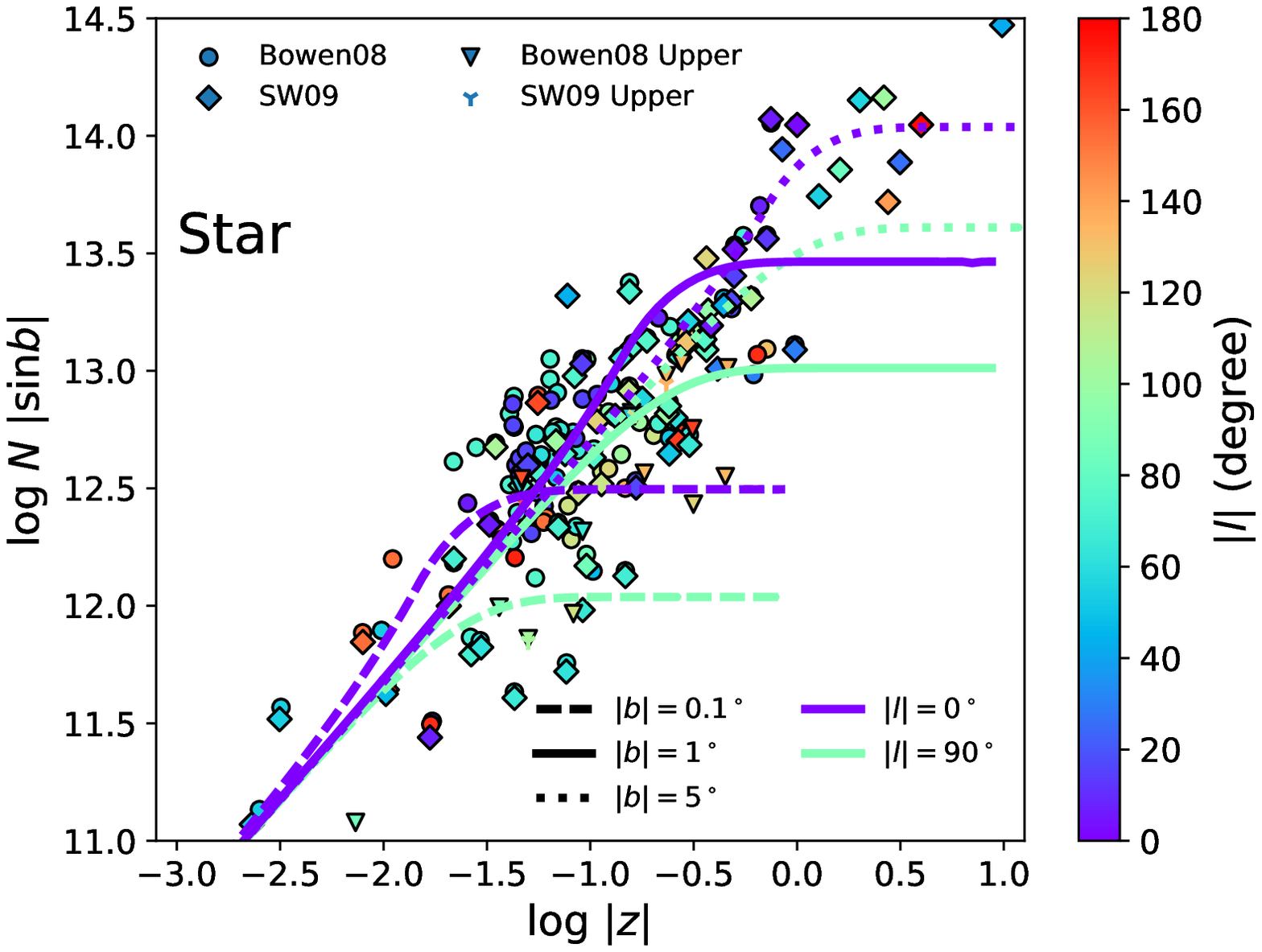}
}
\subfigure{
\includegraphics[width=0.48\textwidth]{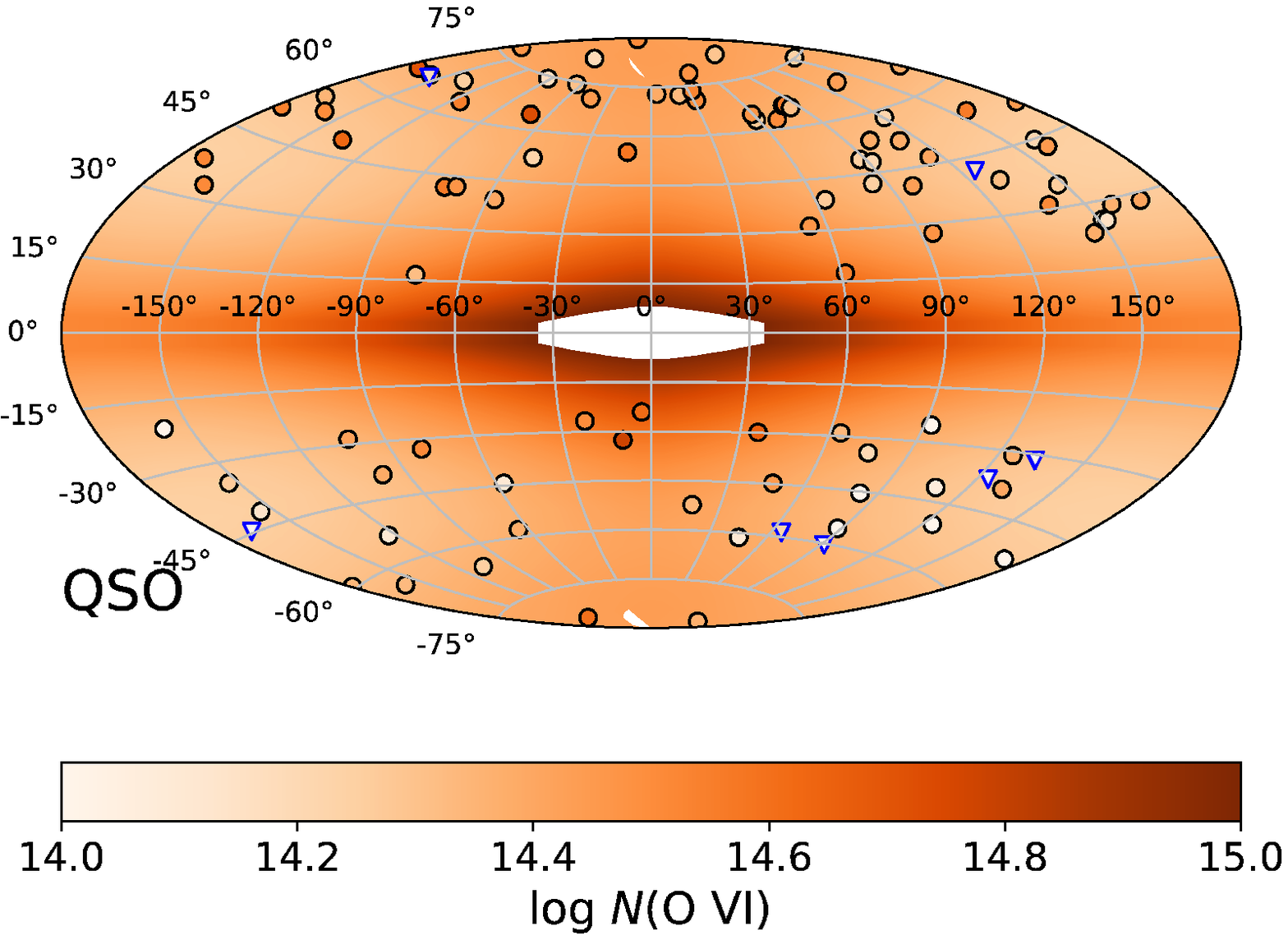}
\includegraphics[width=0.48\textwidth]{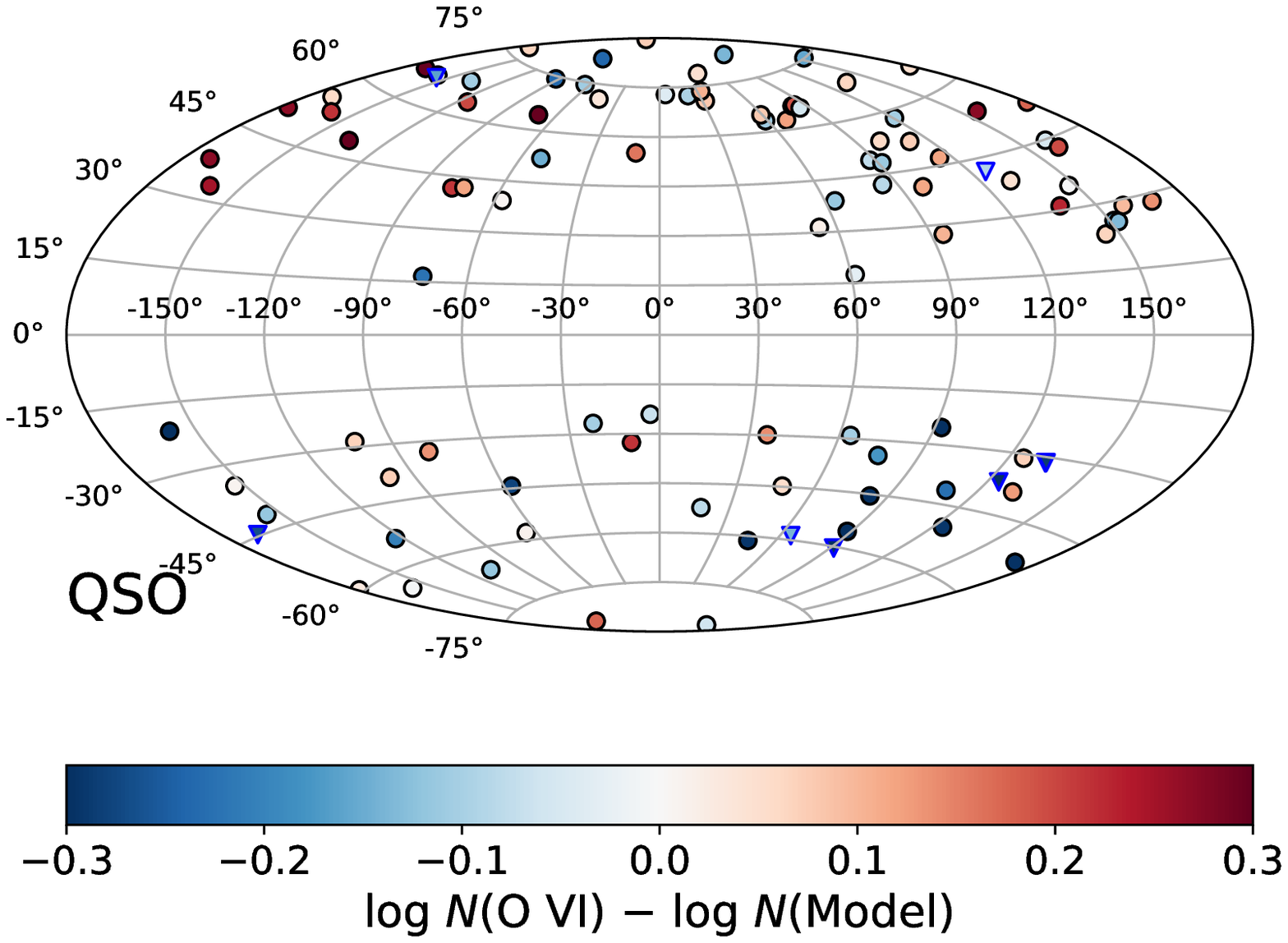}
}
\subfigure{
\includegraphics[width=0.48\textwidth]{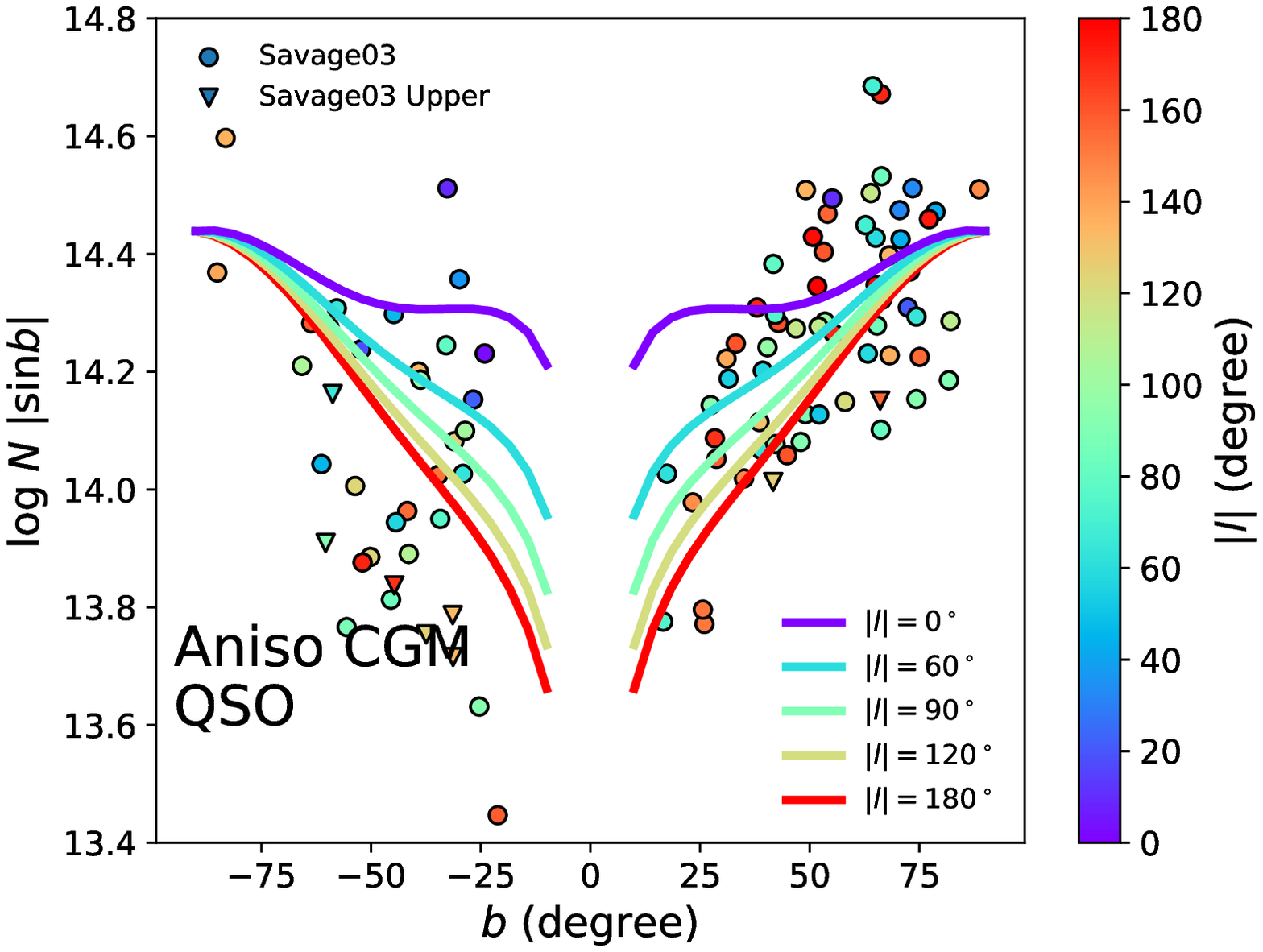}
\includegraphics[width=0.48\textwidth]{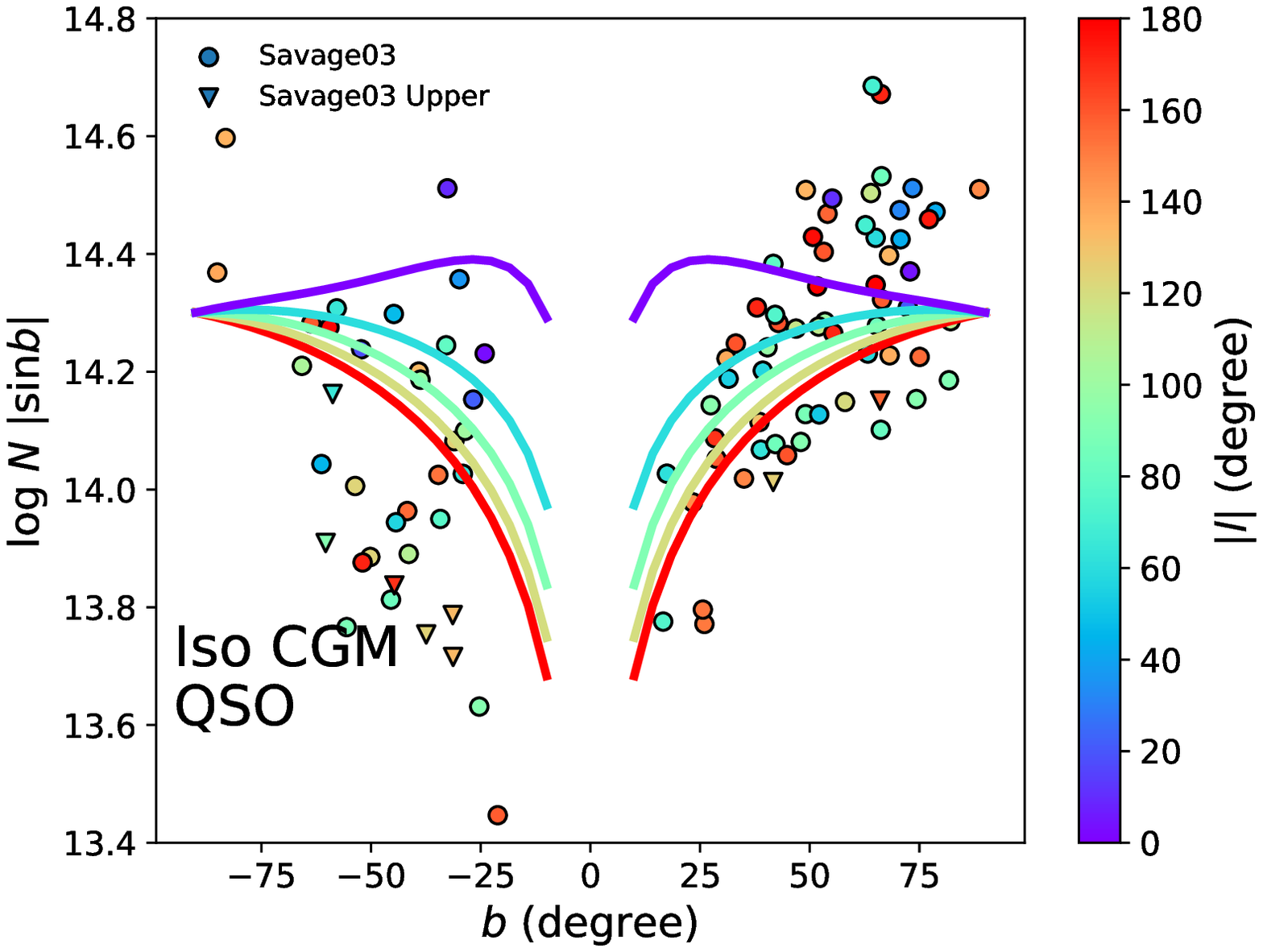}
}
\caption{The comparison between model predictions and observations for \ion{O}{6}. Each panels are the same as Fig. \ref{SiIV_plots}. For the top panels, we also plot the \citet{Bowen:2008aa} sample, which is not included in the fitting. For \ion{O}{6}, the anisotropic CGM model is $4.3\sigma$ better by reducing the total $\chi^2$ of 18.5.}
\label{OVI_plots}
\end{center}
\end{figure*}

For the MW, the azimuthal variation of external galaxies is equivalent to a variation of the CGM column densities as a function of Galactic latitude.
We refine our 2-D disk-CGM model by changing CGM density from an isotropic distribution to an anisotropic distribution with a dependence on Galactic latitude.
In this model, we define two characteristic CGM column densities: the column density along the disk ($N_{\rm mp}$; denoting the mid-plane) and the column density perpendicular to the disk ($N_{\rm nd}$; denoting the normal direction of the disk).
These two directions are similar to the major and minor axis directions for external galaxies.
For simplicity, we assume that the CGM column density depends on Galactic latitude $b$ as an elliptical function:
\begin{equation}
\log N_{\rm CGM}(b) = \sqrt{\log^2 N_{\rm mp} \cos^2 b+ \log^2 N_{\rm nd} \sin^2 b},
\end{equation}
where $N_{\rm mp}$ and $N_{\rm nd}$ are free parameters in our model.
For AGN sightlines, the term $N_{\rm CGM}$ in Equation 2 has a dependence on Galactic latitude ($N_{\rm CGM}(b)$).
{In this model, we assume the variation of the CGM column density is in the logarithmic scale rather than in the linear scale, i.e., $N_{\rm CGM}(b) = (N_{\rm mp}^2 \cos^2 b+ N_{\rm nd}^2 \sin^2 b)^{1/2}$.
In the linear scale variation model, if $N_{\rm nd}$ is much larger than $N_{\rm mp}$ (e.g., a factor of $>3$), the CGM column density will be dominated by $N_{\rm nd} \sin b$, and $N_{\rm mp}$ cannot affect the fitting results.
Therefore, the linear scale variation model does not have the ability to trace the large amplitude CGM variation (i.e., $|\log N_{\rm nd}-\log N_{\rm mp}| > 0.5$ dex).}

The $\chi^2$ fitting results show that the anisotropic CGM model is significantly better than the isotropic CGM model (Table \ref{fits}).
The total $\chi^2$ values are reduced by $15.6-26.4$ and $15.7-21.2$ for \ion{Si}{4} and \ion{O}{6}, respectively.
The mean values of the $\chi^2$ difference are $20.8$ and $18.5$, which lead to a $4.6\sigma$ and $4.3\sigma$ significance considering the degree of freedom (dof) is reduced by 1.
According to the fitting, the CGM column density is higher along the normal direction of the MW disk ($N_{\rm nd}$) than the direction along the disk ($N_{\rm mp}$) by $0.6-0.9$ dex and $\approx 1.0$ dex for \ion{Si}{4} and \ion{O}{6}, respectively.
The differences between $N_{\rm nd}$ and $N_{\rm mp}$ are consistent for different disk density profiles (exponential or Gaussian).
This consistency indicates the CGM anisotropy is a real feature rather than an artificial feature due to the choice of the disk density profiles.
Therefore, we suggest that the anisotropic model is preferred at least at a level of $ 4.0\sigma$ for both \ion{Si}{4} and \ion{O}{6} distributions.
Combining these two ions together, the significance is about $6.3 \sigma$.
However, this result does not imply the CGM has an elliptical geometry, and it is even unknown whether this feature is completely due to the CGM, which will be discussed in Section 4.3.

Adopting the anisotropic CGM model does not affect the disk parameters significantly, but one interesting difference is the larger scale length.
This is the result of the smaller CGM column density at low Galactic latitudes.
In the isotropic CGM model, the CGM column density is dominated by the AGN sample at high Galactic latitudes.
This isotropic column density ($\approx N_{\rm nd}$) is higher than the real column density at low Galactic latitudes ($N_{\rm mp}$), and suppresses the extension of the disk component along the radial direction.
In the fiducial $\rm R_EZ_E$ disk model with an anisotropic CGM, the scale lengths are $5.3\pm1.4$ kpc and $7.8\pm 2.4$ for \ion{Si}{4} and \ion{O}{6}, respectively.
These numbers indicate that the warm gas disk is more extensive than the stellar disk ($\approx 2$ kpc; \citealt{Bovy:2013aa}) and the \ion{H}{1} disk ($\approx 3.5$ kpc; \citealt{Kalberla:2008aa}) at about $2\sigma$.

\begin{figure*}
\begin{center}
\includegraphics[width=0.97\textwidth]{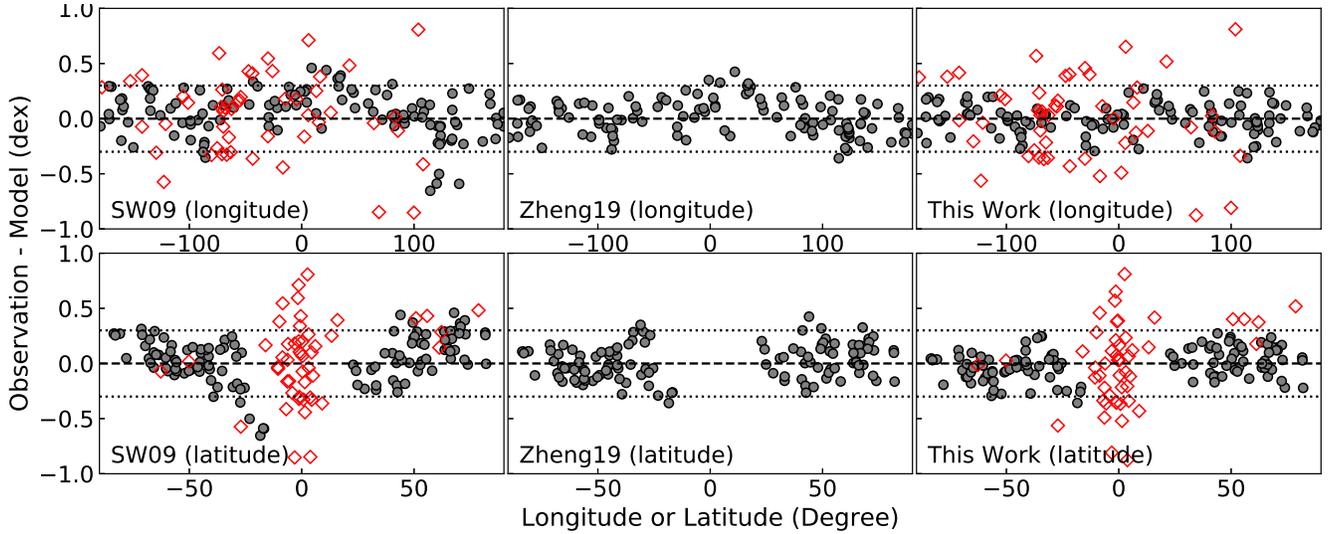}\\
\caption{The comparison of the \ion{Si}{4} residuals between three models: the plane-parallel model (\citetalias{Savage:2009aa}; {\it left panels}), the two-component disk-CGM model (\citetalias{Zheng:2019aa}; {\it middle panels}), and the 2-D disk-CGM model (this work; {\it right panels}). The filled gray circles are the residuals of the AGN sample, while the open red diamonds are the stellar sightlines. Here, we only plot measurements without upper or lower limits. The dotted lines are at levels of $\pm 0.3$ dex. The \citetalias{Savage:2009aa} model are comparable to the 2-D disk model for the stellar sample (flat residuals but large scatters), while the \citetalias{Zheng:2019aa} model cannot reproduce the stellar sample. For the AGN sample, both \citetalias{Savage:2009aa} and \citetalias{Zheng:2019aa} models show hints for unaccounted features, such as the peak around $l=0^\circ$.}
\label{res}
\end{center}
\end{figure*}

Column density predictions by the preferred models ($\rm R_EZ_E$ disk with anisotropic CGM) are compared to the observations in Fig. \ref{SiIV_plots} and Fig. \ref{OVI_plots} for \ion{Si}{4} and \ion{O}{6}, respectively.
We plot the stellar samples in the top panels, which generally follow the model ($\rm R_EZ_E$ disk with isotropic CGM; Fig. \ref{model}) described in Section 3.2.
Different from the plane-parallel slab model \citepalias{Savage:2009aa}, we predict the maximum projected column density has a dependence on Galactic latitude due to the radial profile of the disk (Section 3.2).
However, the difference between low and high Galactic latitude sightlines does not show up for the observed scale heights, since there are no high-$|z|$ stellar sightlines at low latitude ($|b|\lesssim 3^\circ$), which are expected to follow the purple lines in the top left panels.
The observations show the longitude dependence of the projected column density, since the sightlines toward the anti-GC generally have lower projected column density than the sightlines toward the GC.

The middle panels show the global variation of the total column densities predicted for AGN sightlines.
It is of interests to find that the predicted global minimum among the AGN sightlines occurs around $|b|=30^\circ-50^\circ$ around the anti-GC, which is the combination of the disk variation and the anisotropic CGM.
The disk component has the minimum around the polar regions ($|b|\approx 90^\circ$), where has the shortest path length, while the anisotropic CGM component in our model has the minimum along the disk radial direction ($|b|=0^\circ$).
Then, considering these two effects together, the minimum of the total column density will be around $|b|\approx45^\circ$.
Observationally, this feature was found by \citet{Wakker:2012aa}, showing an \ion{O}{6} deficit region at $l = 70-280^\circ$, $b = -60$ to $-10^\circ$.
This deficit is more clear in the southern hemisphere, since the southern hemisphere has systematically lower column densities.

The projected column density of the AGN samples is plotted in the lower panels of Fig. \ref{SiIV_plots} and Fig. \ref{OVI_plots}.
The northern hemisphere sightlines have systematically higher column densities than the southern hemisphere by $0.1-0.2$ dex.
The origin of this north-south asymmetry is beyond the scope of this paper, but we discuss it phenomenologically in Section 4.4.
Here, we do not consider this north-south column density asymmetry in our modeling.
The predicted tendency stated in Section 3.2 matches with the observations, which show the anti-GC sightlines have lower projected column densities.

We find that the anisotropic model can better reproduce the large $\log N$ variation at lower Galactic latitudes for two reasons (Fig. \ref{SiIV_plots} and Fig. \ref{OVI_plots}).
First, the variation of the projected column density at low latitudes is mainly caused by the radial variation of the disk.
The low column density along the disk radial direction ($\log N_{\rm mp}$) allows a more extended disk, which increases the column density variation at low Galactic latitudes, since this variation is due to the variation of the disk (the middle panel of Fig. \ref{model}).
Second, the variation of the CGM column density leads to a steeper decrease of the projected column density at low latitudes around the anti-GC.
Since the CGM column density is higher at $b=90^\circ$, the total column density is also increased at high latitudes.

In all, we prefer the $\rm R_EZ_E$ model with anisotropic CGM profile to other models (Fig. \ref{res}). 
We examine the \ion{Si}{4} column density residuals to evaluate the performance of the plane-parallel slab model (\citetalias{Savage:2009aa}; left column), the two-component disk-CGM model (\citetalias{Zheng:2019aa}; middle column), and our preferred model (right column). 
The plane-parallel slab model (\citetalias{Savage:2009aa}) fits the stellar sample well, which does not have unaccounted features in the residual (i.e., flat residuals over Galactic longitude), although there is a large scatter.
However, there are significant unaccounted structures in the residuals for the AGN sample (low residuals at low Galactic latitudes), although the intrinsic scatter is less than the stellar sample.
The two-component disk-CGM model (\citetalias{Zheng:2019aa}) has comparable residuals to our 2-D disk-CGM model for the AGN sample, but the residuals show a peak around $l=0^{\circ}$. 
Also, this model predicts a disk component of $\log N = 12.1$, which is about one order of magnitude lower than \citetalias{Savage:2009aa} and this work ($\log N\approx 13.3$).
The two-component disk-CGM model does not have distance constraints, so the $\log N$ measurement from stellar sightlines cannot be reproduced in this model \citepalias{Zheng:2019aa}.
Our new model could reproduce the column density measurements from both the stellar and the AGN sightlines equally well without unaccounted features in the residuals.

\section{Discussion}
\subsection{The Comparison with \citetalias{Zheng:2019aa}}
Using the archival data (\citetalias{Savage:2003aa, Savage:2009aa, Zheng:2019aa}), we build a 2-D disk-CGM model to fit the \ion{Si}{4} and \ion{O}{6} column density measurements for warm gas moving at $|v|\lesssim100 \kms$ from both the AGN and stellar sightlines simultaneously.
Previously, the plane-parallel slab model is commonly used to study the MW disk (\citealt{Jenkins:1978aa, Bowen:2008aa}; \citetalias{Savage:2009aa}).
However, the plane-parallel slab model cannot explain a mismatch between the AGN and stellar sample as noted and discussed in \citetalias{Zheng:2019aa}, which is due to the lack of CGM component in the plane-parallel slab model.
\citetalias{Zheng:2019aa} introduced an additional isotropic CGM component to the plane-parallel slab model to account for the CGM contribution in the AGN sightlines.
Here, we mainly compare our 2-D disk-CGM model with the two-component disk-CGM model with 1-D disk \citepalias{Zheng:2019aa}.

For the model setting, the major difference is the inclusion of the disk radial distribution in our 2-D disk-CGM model (Section 3.2), while a minor difference is an improvement from the isotropic CGM model to the Galactic latitude-dependent CGM model (Section 3.3).
In \citetalias{Zheng:2019aa}, the disk component is still the 1-D plane-parallel slab model, which leads to a lower disk component.
This is because the plane-parallel slab model has a constant projected column density of AGN over different Galactic latitudes.
Therefore, the low Galactic latitude sightlines with low projected column densities (Fig. \ref{SiIV_plots} and Fig. \ref{OVI_plots}) lead to a low value of the disk component in the two-component disk-CGM model \citepalias{Zheng:2019aa}.
For the CGM component, we find that the column density distribution of MW CGM is likely to be a function of Galactic latitude instead of an isotropic one (Section 3.3). 
Since the AGN samples are mostly around the high latitude ($|b| \gtrsim 30^\circ$), this CGM modification lead to a significant difference ($\approx0.7-1.0$ dex for both \ion{Si}{4} and \ion{O}{6}) at low Galactic latitudes between our models and the two-component disk-CGM model \citepalias{Zheng:2019aa}.

Another difference is the adopted statistical method, where we used the $\chi^2$ optimization, while \citetalias{Zheng:2019aa} used the Bayesian frame.
Assuming the Gaussian distribution for the measurement uncertainty and the uniform prior, these two methods are equivalent in the sense to obtain the minimum of $\chi^2$ or the maximum of likelihood.
Besides the method to obtain the fitting results, another difference in the statistical method is the choice of the likelihood or the uncertainty distribution.
\citetalias{Zheng:2019aa} assumed the column density uncertainty follows a normal distribution (the linear scale) rather than a lognormal distribution, while the latter distribution is adopted in our models.
Although the real distribution of the uncertainty is unknown, the lognormal distribution is more used for the data with a large variation (e.g., one order of magnitude).
It is worth to notice that these two distributions are similar to each other when the uncertainty is small ($\approx 0.02$ for most sightlines in \citetalias{Zheng:2019aa}).

\citetalias{Zheng:2019aa} also used the block bootstrapping to account for the possible unknown large scale structures.
From the residual map in Fig. \ref{SiIV_plots} and Fig. \ref{OVI_plots}, we noticed the north-south asymmetry, which is the most prominent variation over the entire sky.
This feature is addressed and discussed phenomenologically in Section 4.4.
In section 4.5, we introduce several blocking tests for large scale structures, which show consistent results with the unblocked fitting.

Another minor difference is that we obtain the patchiness parameter by reducing the reduced $\chi^2$ to 1, while \citetalias{Zheng:2019aa} implemented the patchiness parameter in the Bayesian frame.
Again, \citetalias{Zheng:2019aa} assumed a normal distribution rather than a lognormal distribution for this intrinsic scatter.
Therefore, their patchiness parameter estimate is $N_{\rm p} = 1.4\times 10^{13}$ in the linear scale.
We convert it into the logarithmic scale by $\log_{10}e \times  \frac{N_{\rm p}}{N_{\rm SiIV}}\approx 0.179$, where $N_{\rm SiIV}$ is the mean column density of the AGN sample of \ion{Si}{4}.
This value is larger than the one in our models (0.13 dex; Table \ref{sample}), and we suggest that this difference is mainly due to the inclusion of the disk radial profile to better account for the column density scatters as seen in the AGN data.

\begin{table*}
\begin{center}
\caption{The Joint Fitting Results of \ion{Si}{4} and \ion{O}{6}}
\label{joint_fits}
\begin{tabular}{lccccccccc}
\hline
\hline
Ion & $\log n_0$ & $r_0$ & $z_0$ & ${\Delta\log N^{\rm CGM} } ~^a$ & $\log N_{\rm nd}^{\rm CGM}$ & red. $\chi^2$ (dof) & $\log n_{\odot}^{\rm disk}$ & $\log n_\odot^{\rm disk} z_0$ & $\log M_{\rm disk}$\\
 & $(\cc)$ & kpc & kpc & $\rm dex~ (\cmsq)$ & $(\cmsq)$ & & $(\cc)$ & $(\cmsq)$ & $(M_\odot)$\\
 \hline
 \ion{Si}{4} & $-8.02\pm0.12$ & $6.1 \pm 1.2$ & $2.6 \pm 0.4$ & $0.82\pm0.32$ & $13.32 \pm 0.07$ & 1.126 (376) & -8.63 & 13.28 & 3.91 \\
\ion{O}{6} & $-7.22\pm0.12$ & $...$ & $...$ & $...$ & $14.17 \pm 0.08$ & $...$ & -7.83 & 14.08 & 4.46 \\
 
\ion{Si}{4} & $-8.04\pm0.13$ & $6.1 \pm 1.2$ & $2.9 \pm 0.5$ & $0.86\pm0.43$ & $13.30 \pm 0.08$ & 1.123 (374) & -8.64 & 13.30 & 3.93 \\
\ion{O}{6} & $-7.21\pm0.12$ & $...$ & $2.3\pm 0.5$ & $0.74 \pm 0.32$ & $14.20 \pm 0.08$ & $...$ & -7.81 & 14.04 & 4.42 \\

\ion{Si}{4} & $-7.92\pm0.16$ & $5.2 \pm 1.2$ & $2.6 \pm 0.4$ & $0.69\pm0.32$ & $13.32 \pm 0.07$ & 1.117 (374) & -8.64 & 13.26 & 3.86 \\
\ion{O}{6} & $-7.36\pm0.13$ & $8.0 \pm 2.3$ & $...$ & $0.99 \pm 0.49$ & $14.19 \pm 0.08$ & $...$ & -7.82 & 14.08 & 4.56 \\

\ion{Si}{4} & $-7.96\pm0.16$ & $5.5 \pm 1.2$ & $2.7 \pm 0.5$ & $0.81\pm0.32$ & $13.32 \pm 0.08$ & 1.118 (374) & -8.63 & 13.29 & 3.89 \\
\ion{O}{6} & $-7.32\pm0.15$ & $7.5 \pm 2.1$ & $2.4 \pm 0.5$ & $...$ & $14.18 \pm 0.08$ & $...$ & -7.82 & 14.06 & 4.50 \\
\hline
\end{tabular}
\end{center}
Notes: Every two lines are one model, since these are joint models for both \ion{Si}{4} and \ion{O}{6}. The blank parameters of \ion{O}{6} are tied to \ion{Si}{4}.\\
{$^a$ $\Delta\log N^{\rm CGM}  = \log N_{\rm nd}^{\rm CGM} - \log N_{\rm mp}^{\rm CGM}$. Positive values indicate that the CGM column density is higher in the direction perpendicular to the disk $\log N_{\rm nd}^{\rm CGM}$ than the radial direction ($\log N_{\rm mp}^{\rm CGM}$).}
\end{table*}

\subsection{The Warm Gas Disk}
The warm gas at $(1 - 5) \times 10^5$ K is important for gas assembly and recycling in a galaxy due to their high cooling rates and short lifetimes ($\approx 10$ Myr; \citealt{Oppenheimer:2013aa}).
Theoretically, this gas normally trace the interaction layer between cool and hot gases, and the cooling from hotter mediums \citep{Gnat:2010aa, Kwak:2015aa}. 
These phenomena are usually associated with galactic outflows (feedback processes), infall gas (gas accretion), and interactions between the disk and the CGM \citep{McQuinn:2018aa, Qu:2018aa}.
Therefore, one could obtain unique insights into the disk and the CGM formation by observing the warm gas. 

The scale height is a key property of the warm gas disk, since it indicates how extensive the disk is, which is a test for the ionization mechanism and the gas origin (\citealt{Bowen:2008aa}; \citetalias{Savage:2009aa}; \citealt{Wakker:2012aa}).
For example, the scale height should be larger for ions with higher ionization potentials under collisional ionization equilibrium (CIE).
However, as shown in \citetalias{Savage:2009aa}, the \ion{O}{6} disk ($z_0 = 2.6\pm0.5$ kpc) has slightly lower scale height than both \ion{Si}{4} ($3.2_{-0.6}^{+1.0}$ kpc) and \ion{C}{4} ($3.6_{-0.8}^{+1.0}$ kpc) in the \citetalias{Savage:2009aa} model.
This phenomenon might indicate that the Galactic \ion{Si}{4} and \ion{O}{6} are produced under different ionization mechanisms \citepalias{Savage:2009aa}.

However, as stated in Section 3.2, we find that the \ion{Si}{4} scale height is reduced from $3.2_{-0.6}^{+1.0}$ kpc \citepalias{Savage:2009aa} to $2.6\pm0.6$ kpc (the $\rm R_EZ_E$ model with the anisotropic CGM).
The \ion{O}{6} scale height ($2.6\pm0.6$ kpc) is similar to \citetalias{Savage:2009aa} ($2.6 \pm 0.5$ kpc).
Therefore, our models do not support that \ion{Si}{4} and \ion{O}{6} have different scale heights.
The different behaviors between \ion{Si}{4} and \ion{O}{6} scale heights are because of: the inclusion of the disk radial profile and the anisotropic CGM component, the different samples, and the exclusion of AGN sightlines around the north Galactic polar region in \citetalias{Savage:2009aa}. 

Besides the scale heights, we find that the scale lengths are also similar between \ion{Si}{4} and \ion{O}{6} within $1 \sigma$.
Therefore, we consider whether both of the ions follow the same density profile distributions of the disk component.
A joint fitting model is applied to \ion{Si}{4} and \ion{O}{6} samples simultaneously, where we tie the parameters of the \ion{O}{6} model to the \ion{Si}{4} model, including the scale length ($r_0$), the scale height ($z_0$), and the CGM difference between two axes ($\Delta \log N^{\rm CGM}$).
In the model where the three parameters are all tied (Table \ref{joint_fits}), the difference of total $\chi^2$ is 3.51 compared to the best model with all parameters are free (the models in Table \ref{fits}).
Because the best model has a 3 more dof, it is $1\sigma$ better than the most limited model, so the best model is not a significantly better model.
We also tie these three parameters in turn to check which is the most dominant factor in the $\chi^2$ difference.
We find that tying the scale length leads to the highest $\chi^2$, but the difference is still insignificant.
Therefore, we prefer the most limited model with all three parameters tied, and suggest that there is no significant difference of the density profile between \ion{Si}{4} and \ion{O}{6} adopting the new models.

Our model measures the scale length of the warm gas disk of the MW for the first time.
We can further estimate the total mass of the warm gas disk of the MW. 
First, we obtain {the total number of ions ($\mathcal{N}_{\rm total}^{\rm disk}$; for \ion{Si}{4} or \ion{O}{6})} within the warm gas disk by integrating the ion number density over the radial and vertical directions:
\begin{equation}
\mathcal{N}_{\rm total}^{\rm disk} = n_0 \int_0^{R_{\rm vir}} {\rm d} r \exp (-\frac{r}{r_0}) \int_{-R_{\rm vir}}^{R_{\rm vir}} {\rm d} z \exp (-\frac{|z|}{z_0}).
\end{equation}
Then, we calculate the masses of \ion{Si}{4} and \ion{O}{6} ions in the warm gas disk for different models (Table \ref{fits} and Table \ref{joint_fits}).
For each ion, various models lead to similar ion masses within 0.3 dex.
Summarizing our results, we obtain the mass of \ion{Si}{4} is $\log (M/M_\odot) = 3.8 \pm 0.1$, while the \ion{O}{6} mass is $\log (M/M_\odot) = 4.4 \pm 0.2$.
To obtain the total mass of the \ion{Si}{4} or \ion{O}{6}-bearing gases, we assume the warm gas has the solar metallicity, and adopt log(Si/H) and log(O/H) solar abundance values from \citet{Asplund:2009aa}.  
Also, we assume the average ionization fraction of $0.2$ and $0.1$ for \ion{Si}{4} and \ion{O}{6}, respectively (about half of the maximum in CIE or PIE {to represent the average ionization fraction}; \citealt{Gnat:2007aa, Oppenheimer:2013aa}).
Then, {the expected total number of hydrogen atoms} is $\mathcal{N}_{\rm H} = \mathcal{N}_{\rm total}^{\rm disk}/f/a$, where $f$ is the ionization fraction of \ion{Si}{4} or \ion{O}{6}, and $a$ is the abundance of silicon or oxygen.
Taking the helium mass into account, the total mass of the warm gas disk is $1.3 \mathcal{N}_{\rm H} m_{\rm H}$, where $m_{\rm H}$ is the hydrogen atom mass.
Finally, the derived total masses of the warm gas disk based on \ion{Si}{4} and \ion{O}{6} are
\begin{eqnarray}
\log (M_{\rm H}/M_\odot)_{\rm SiIV}^{\rm disk} = (7.6 \pm 0.1) - \log \frac{f_{\rm SiIV}}{0.2} - \log  \frac{Z}{Z_\odot}, \notag \\
\log (M_{\rm H}/M_\odot)_{\rm OVI}^{\rm disk} = (7.6 \pm 0.2) - \log \frac{f_{\rm OVI}}{0.1} - \log \frac{Z}{Z_\odot},
\end{eqnarray}
which are similar to each other.

The similarities of shapes and masses between \ion{Si}{4} and \ion{O}{6} disks indicate that these two ions might trace the same gases.
However, it does not mean that these two ions are cospatial, since the ion ratio (\ion{Si}{4}/\ion{O}{6}) shows large scatters ($\approx 0.5$ dex; \citetalias{Savage:2009aa}).
\ion{Si}{4} and \ion{O}{6} occupy the same space at large-scale (Galactic scale) due to the similarities of the disk shapes, but these two ion-bearing gases are clumpy to be non-cospatial at small-scale (single cloud size; kpc size; \citealt{Werk:2019aa}).
The \ion{Si}{4} gas is more clumpy than \ion{O}{6} because it has larger intrinsic scatters (the patchiness parameter; Table \ref{sample}).
The same shapes of the \ion{Si}{4} and the \ion{O}{6} disk profiles from our models indicate that the warm gas disk cannot be in equilibrium.
If these ions are in photoionization equilibrium, the \ion{Si}{4} gas should have a larger scale height, while the thermal-supported collisional disk predicts the opposite behavior.

A possible explanation of the same scale heights for \ion{Si}{4} and \ion{O}{6} is that these ions are produced by feedback processes (e.g., the Galactic fountain; \citealt{Bregman:1980aa, Melso:2019aa}).
In the Galactic fountain, the gas could be IVCs, which are separate clouds \citep{Wakker:2008aa, Shull:2009aa, Werk:2019aa}.
Then, the \ion{Si}{4} gas is close to the core of \ion{H}{1}, while the \ion{O}{6} gas is likely to be the envelope, since \ion{Si}{4} has a lower excitation potential.
The scale heights of these two ions are both set by the ejection due to Galactic feedback.
It is also explained that the \ion{Si}{4} gas is more clumpy than the \ion{O}{6} gas, since as an envelope, the \ion{O}{6} gas should have a larger volume filling factor.

As a comparison to the neutral gas, the \ion{H}{1} disk has a total mass of $7.1\times 10^9 ~M_\odot$, a scale height of $0.15$ kpc, and a scale length of $3.25$ kpc within 30 kpc \citep{Kalberla:2008aa, Nakanishi:2016aa}.
Besides the thin \ion{H}{1} disk component, there is also a more extensive \ion{H}{1} disk with a scale height of $1.6_{-0.4}^{+0.6}$ kpc, which contains a mass of $3.2_{-0.9}^{+1.0} \times 10^8~M_\odot$ (named as the \ion{H}{1} halo in \citealt{Marasco:2011aa}).
The warm gaseous disk has a larger scale height than the thick \ion{H}{1} disk, while the mass is about one order of magnitude lower.

\subsection{The Anisotropic CGM}
As stated in Section 3.2, the preferred CGM component in our model is anisotropic with a dependence on Galactic latitude.
The joint fitting of \ion{Si}{4} and \ion{O}{6} shows that there is an enhancement of $\Delta \log N = 0.82\pm0.32$ for the column density perpendicular to the disk compared to the direction along the disk.
It is worth noticing that although the component is named as ``CGM", it does not mean that this enhancement of the column density is completely due to the CGM of the MW.
{This column density enhancement could be due to the enriched CGM of the MW or the interaction layer between the disk and the CGM (e.g., interface layers around low-intermediate velocity clouds).}

In the first scenario, the CGM above the disk is enriched by feedback processes from the disk which ejected (and recycled) materials/metals into the CGM.
Also, the escaping ionizing fluxes are more intense in the $z$-direction, which could lead to higher ionization states (\ion{Si}{4} and \ion{O}{6}) by photoionization.
{Another possibility is that there is an interaction layer between the disk and the halo gas above the disk, such as the Galactic fountain \citep{Bregman:1980aa}, which can be observed as low-intermediate velocity clouds \citep{Wakker:2008aa, Werk:2019aa}.}
If this component cannot be included in the disk component in our modeling, then it has to be attributed to the anisotropic ``CGM" component, which might be the case here.
Although these two possibilities are both associated with the feedback processes, the difference is the location of the gases, which could affect the estimation of the mass of the MW warm CGM.
However, current observations cannot determine the location of these gases.
Hereafter, we assume it is the enriched CGM scenario.

{To estimate the mass of the warm CGM, we calculate the the average CGM column density over the entire sky, which is $\frac{1}{2}\int_{-\pi/2}^{\pi/2} {\rm d} b N_{\rm CGM}(b) \cos b$.
The average CGM column densities are $\log N = 12.84$ and $\log N = 13.70$ for \ion{Si}{4} and \ion{O}{6}, respectively.}
The maximum radius of the CGM is fixed as the virial radius of the MW (250 kpc).
The total ion mass is $\log (M/M_\odot) = (5.6 \pm 0.2) + 2\log (R_{\rm max}/250\rm~kpc)$ assuming the uniform density distribution for \ion{Si}{4}, and $\log (M/M_\odot) = (6.3 \pm 0.2) + 2\log (R_{\rm max}/250\rm~kpc)$ for \ion{O}{6}, where $R_{\rm max}$ is the maximum radius of the CGM. 
The metallicity of the MW CGM is assumed to be $0.5~Z_\odot$ \citep{Bregman:2018aa}, and the average ionization fraction is about the half of the peak from CIE or PIE (similar to the disk calculation in Section 4.2).
Then, the \ion{Si}{4} and \ion{O}{6}-bearing gases have masses of
\begin{eqnarray}
\log (M_{\rm H}/M_\odot)_{\rm SiIV}^{\rm CGM} & =& (9.8\pm 0.2) - \log \frac{f_{\rm SiIV}}{0.2} - \log \frac{Z}{0.5Z_\odot} , \notag \\
& & + 2 \log \frac{R_{\rm max}}{250 \rm kpc}, \notag \\
\log (M_{\rm H}/M_\odot)_{\rm OVI}^{\rm CGM} &=& (9.8\pm 0.2) - \log \frac{f_{\rm OVI}}{0.1} - \log \frac{Z}{0.5Z_\odot} , \notag \\
& & + 2 \log \frac{R_{\rm max}}{250 \rm kpc}.
\end{eqnarray}
Different from the disk component, we cannot constrain whether \ion{Si}{4} and \ion{O}{6} have similar shapes for the CGM, but if we assume they follow the same density profile, the masses are the same for these two ions.
If one wants to estimate the mass for the interaction-layer scenario, one could use $\log N = 12.50$ and $\log N =13.36$ instead in the mass estimation, which is a difference of $\approx 0.34$ dex.

\begin{table*}
\begin{center}
\caption{The Fitting Results of the North-South Asymmetry}
\label{NS_fits}
\begin{tabular}{lcccccccccc}
\hline
\hline
Ion & $\log n_0$ $^a$& $\Delta \log n_0^{\rm NS}$ $^b$ & $r_0$ & $z_0^{\rm N}$ & $z_0^{\rm S}$ & $\Delta\log N^{\rm CGM}$ & $\log N_{\rm nd}^{\rm CGM}$ $^a$ & $\Delta\log N_{\rm CGM}^{\rm NS}$ $^b$ & red. $\chi^2$ (dof) \\
 & $(\cc)$ & $(\cc)$ & kpc & kpc & kpc & $\rm dex~ (\cmsq)$ & $(\cmsq)$ & $(\cmsq)$ \\
\hline
\ion{Si}{4} & $-8.01\pm 0.12$ & $0.06\pm0.05$ & $6.0\pm1.1$ & $3.4\pm0.6$ & $2.5\pm0.4$ & $0.92\pm0.46$ & $13.21\pm0.14$ & $-0.09\pm0.15$ & 1.034 (373)\\
\ion{O}{6} & $-7.20\pm 0.12$ & $...$ & $...$ & $...$ & $...$ & $...$ & $13.96\pm0.16$ & $...$ & $...$\\
\ion{Si}{4} & $-7.97\pm 0.12$ & $0.14\pm0.02$ & $6.0\pm1.0$ & $2.9\pm0.4$ & $...$ & $0.98\pm0.46$ & $13.26\pm0.08$ & $...$ & 1.045 (375)\\
\ion{O}{6} & $-7.17\pm 0.11$ & $...$ & $...$ & $...$ & $...$ & $...$ & $14.06\pm0.11$ & $...$ & $...$\\
\ion{Si}{4} & $-8.03\pm 0.12$ & $...$ & $5.9\pm1.1$ & $3.5\pm0.5$ & $2.3\pm0.4$ & $0.83\pm0.41$ & $13.26\pm0.08$ & $...$ & 1.039 (375)\\
\ion{O}{6} & $-7.23\pm 0.12$ & $...$ & $...$ & $...$ & $...$ & $...$ & $14.04\pm0.11$ & $...$ & $...$\\
\ion{Si}{4} & $-7.95\pm 0.13$ & $...$ & $5.2\pm1.0$ & $2.5\pm0.4$ & $...$ & $0.56\pm0.20$ & $13.40\pm0.06$ & $0.29\pm0.09$ & 1.063 (375)\\
\ion{O}{6} & $-7.18\pm 0.12$ & $...$ & $...$ & $...$ & $...$ & $...$ & $14.25\pm0.06$ & $...$ & $...$\\
\hline
\end{tabular}
\end{center}
Notes: Every two lines are one model, since these are joint models for both \ion{Si}{4} and \ion{O}{6}. The blank parameters of \ion{O}{6} are tied to \ion{Si}{4}.\\
$^a$ The disk density and the CGM column density are for the northern hemisphere.\\
$^b$ For the difference between two hemispheres, the positive value indicates that the northern hemisphere is higher than the southern hemisphere.
\end{table*}

The mass of the CGM has a dependence on the radial profile of the density.
Although the radial profile of the warm gas cannot be determined using current observations for the MW, we show the effect of this variation as the following.
For simplicity, we assume a $\beta$-model of $n(r) = n_0 r^{-3\beta}$ (a power law model), which is empirical for the MW hot gas \citep{Li:2017aa} and warm gas in external galaxies \citep{Werk:2013aa, Johnson:2015aa}.
Then, the new CGM mass is 
\begin{equation}
\frac{M_{\beta}}{M_{\rm u}} = \frac{1-3\beta}{1-\beta} \frac{R_{\rm max} - R_{\rm min}}{R_{\rm max}^3- R_{\rm min}^3} \frac{R_{\rm max}^{3-3\beta} - R_{\rm min}^{3-3\beta}}{R_{\rm max}^{1-3\beta}- R_{\rm min}^{1-3\beta}},
\end{equation}
where $M_{\beta}$ is the mass in the $\beta$-model, while $M_{\rm u}$ is the mass in the uniform density model (Equation 5).
$R_{\rm max}$ and $R_{\rm min}$ are the maximum and the minimum radii.
With a boundary of 10 kpc and 250 kpc, the ${M_{\beta}}/{M_{\rm u}}$ ratio is 0.24 with $\beta = 1/2$ (the theoretical hydrostatic equilibrium solution; \citealt{Mo:2010aa}), which is a correction of $-0.6$ dex for Equation 6.
Generally, a larger $\beta$ leads to a smaller mass of the CGM.
Varying $\beta$ from $1/3$ to $2/3$, the mass ratio varies from $0.45$ to $0.12$, and the mass is always lower than the uniform model.
Therefore, we suggest that the mass in the Equation 6 is the upper limit if the radial profiles of \ion{Si}{4} and \ion{O}{6} are decreasing at larger radii with the same assumptions of the abundance and the ionization fraction.
The suggested mass region is $\log (M/M_\odot) \approx 8.9-9.5$ with a correction of $- \log \frac{Z}{0.5Z_\odot}$.

The mass of the HVCs are not included in the previous discussion, since the \citetalias{Savage:2003aa} and \citetalias{Zheng:2019aa} samples only measured absorption features at low and intermediate velocities ($|v|\lesssim 100\kms$).
One of the major contributor of the HVCs is the Magellanic Systems (MS), which has a total mass of $\log (M/M_\odot) \approx 9.3$ for atomic and warm-ionized gases: $\approx 4.9\times 10^8~ M_\odot$ in \ion{H}{1}; $\approx1.0\times10^9~M_\odot$ in the warm gas in MS; and $\approx 5.5 \times 10^8~M_\odot$ in the envelope of the MS \citep{Bruns:2005aa, Fox:2014aa}.
Besides the Magellanic systems, other HVCs have a total \ion{H}{1} mass of $2.6\times 10^7~M_\odot$ \citep{Wakker:2004aa, Putman:2012aa}.
{Assuming other HVCs have a similar the \ion{H}{1}/total warm gas ratio of MS (1:4), the total mass of other HVCs is about $1 \times10^8 ~M_\odot$ \citep{Lehner:2012aa}.}
Then, the total HVC mass in the MW is $\log (M/M_\odot) \approx 9.4$, which is comparable to the derived mass of low- and intermediate-velocity gas in this paper.
Therefore, the total mass of the warm-ionized gaseous halo is about $\log (M/M_\odot) \approx 9.5-9.8$ for the all velocity range.

This derived mass is consistent with the mass of $\log (M/M_\odot) \gtrsim 9.3$ reported in \citetalias{Zheng:2019aa}, which only used the \ion{Si}{4} AGN sample to estimate the CGM column density of the MW.
Our estimation of the warm CGM mass is comparable to the Andromeda galaxy, which has a total mass of $\log (M/M_\odot) \approx 9.1 - \log (Z/Z_\odot)$ for the warm gas (up to \ion{C}{4}; \citealt{Lehner:2015aa}).
The warm CGM mass of the MW is consistent with some $L^*$ galaxy samples at redshifts of $z\lesssim 0.2$ with $\log (M/M_\odot)\approx9.5-10.4$ \citep{Stocke:2013aa}, while there are also samples showing significant differences of $L^*$ galaxies at $z\approx 0.2$ (e.g., COS-Halos), which obtained a mass of $\log (M/M_\odot)\approx 10.8-11.0$ \citep{Werk:2014aa, Prochaska:2017aa}.
However, significantly different masses are derived with different models using the same COS-Halos data, such as $\log (M/M_\odot) \approx 10.1$ \citep{Stern:2016aa, Bregman:2018aa}.
These uncertainties suggest that the mass estimation of CGM is model-depended, but the local $L^*$galaxies (i.e., the MW and the Andromeda) do not favor a warm CGM with a mass comparable to the stellar mass.

\subsection{Comments on the North-South Asymmetry}
As shown in Fig. \ref{SiIV_plots} and Fig. \ref{OVI_plots}, there is a significant north-south (NS) asymmetry for the observed scale height of AGN samples, which indicates the asymmetry of the warm gas distribution.
This asymmetry is similar to the NS asymmetry of the Galactic X-ray background, which shows more soft X-ray emission in the northern hemisphere \citep{Snowden:1997aa}.
The physical origin of this asymmetry is unclear, and beyond the scope of this paper.
However, it is of interest to determine the origin of the warm gas asymmetry phenomenologically (i.e., the disk or the CGM).

Based on previous results, we assume \ion{Si}{4} and \ion{O}{6} have the same behaviors for both the disk and the CGM: the scale length, the scale height, and the CGM difference between two axes.
Considering the NS asymmetry, there are three possible variations between two hemispheres: the scale height of the disk; the disk normalization density; and the CGM column density.
Here, we ignore the possible difference of the scale length, which is fixed to the same for both hemispheres.
Then, this model can have different scale heights and different disk density normalizations between the north and south disks, and different azimuthal CGM column densities (Table \ref{NS_fits}).
The total $\chi^2$ difference is 35.1, which is $5.3 \sigma$ with a dof difference of 3.
The fitting reveals that the differences of the disk density normalizations and the CGM column density normalizations are close to zero, within the uncertainty.
The largest variation is due to the difference in the scale heights.

\begin{figure*}
\begin{center}
\subfigure{
\includegraphics[width=0.48\textwidth]{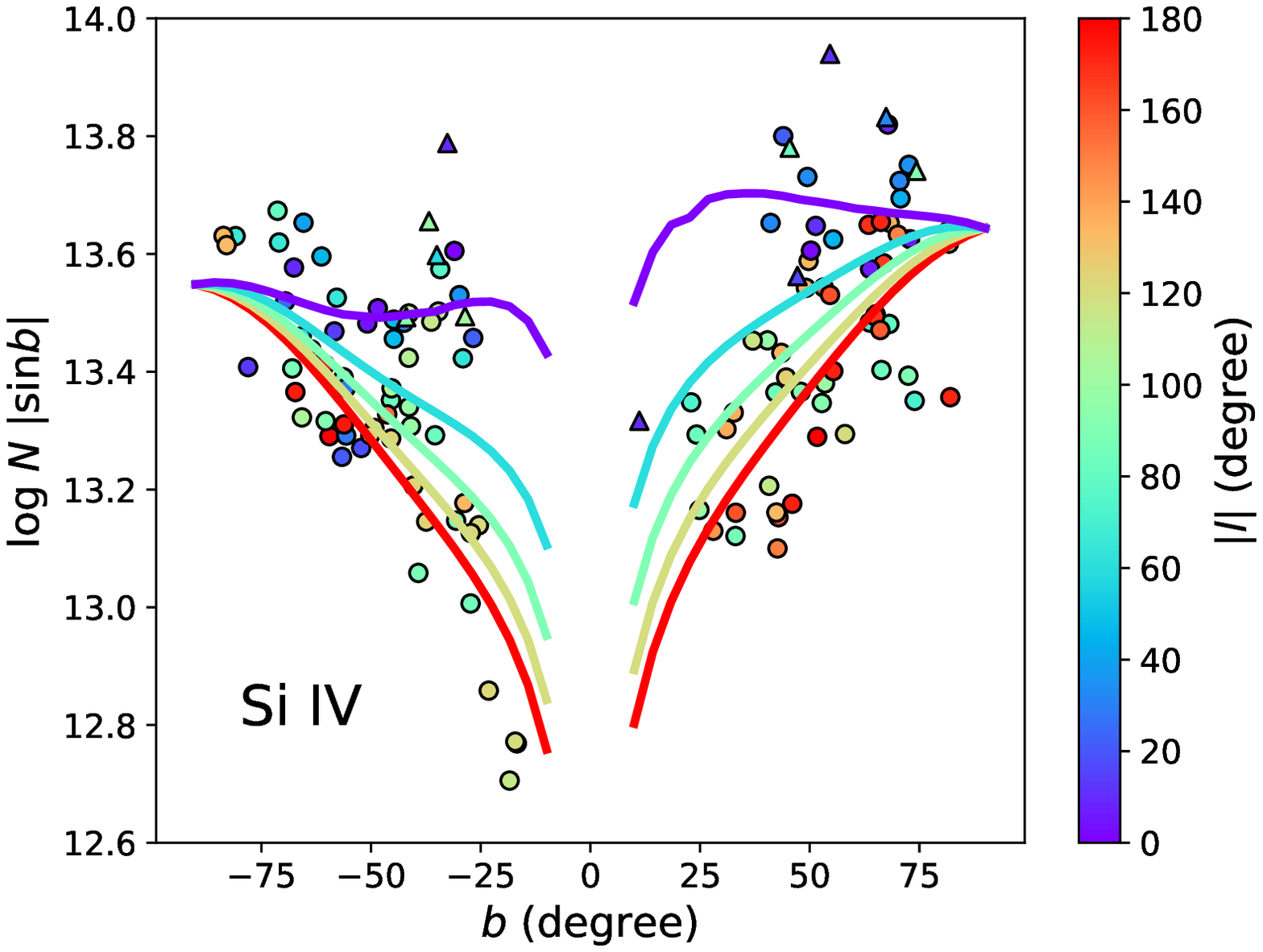}
\includegraphics[width=0.48\textwidth]{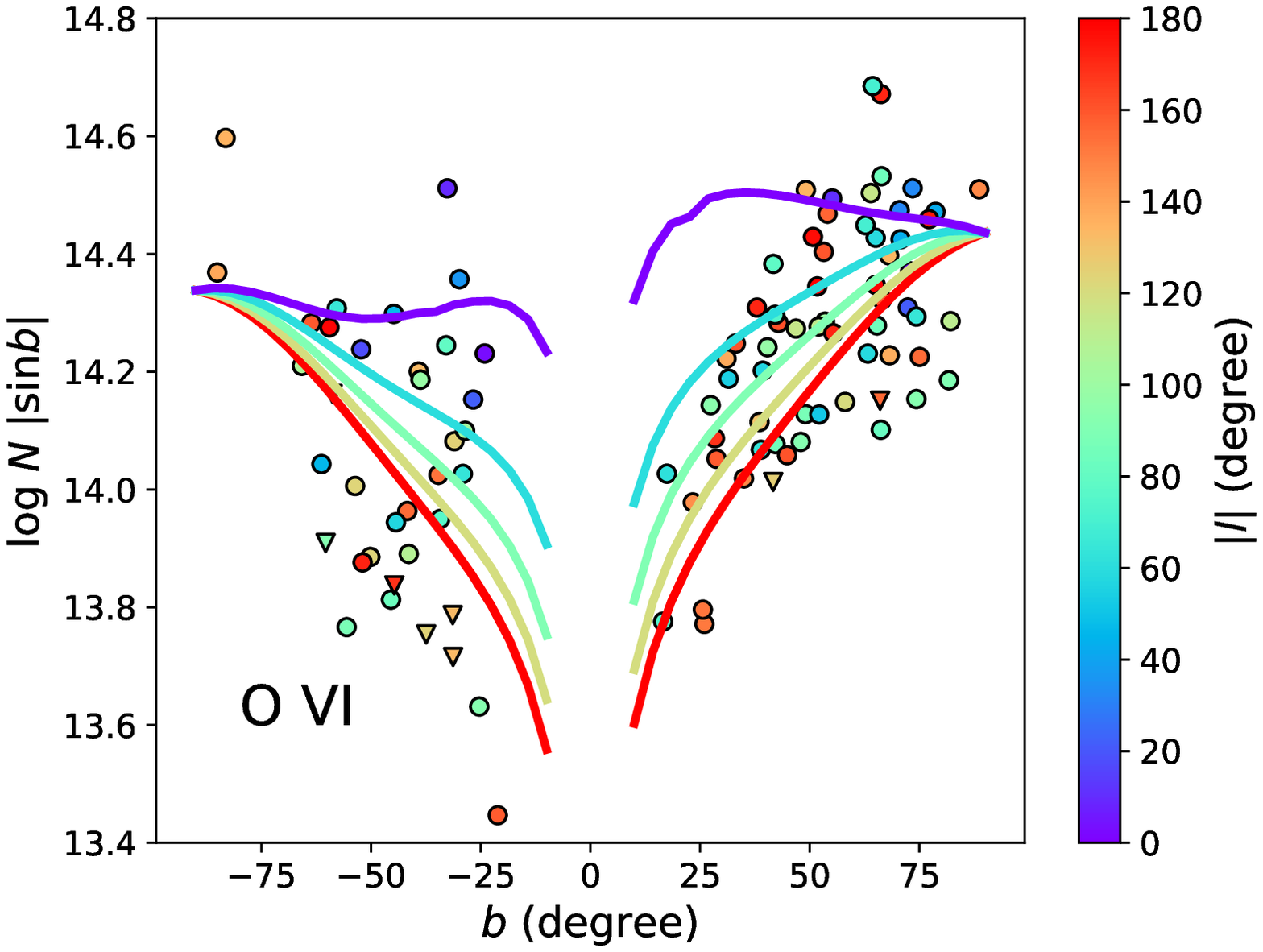}
}
\caption{The varied scale height model to account for the north-south asymmetry. The data and model are color encoded in the same way as Fig. \ref{SiIV_plots} and Fig. \ref{OVI_plots}.}
\label{NS_sh}
\end{center}
\end{figure*}

Quantitatively, we vary these parameters individually to determine the dominant factor (Table \ref{NS_fits}).
The fitting results show that the scale height is the dominant parameter rather than the disk normalization or the CGM normalization with the smallest reduced $\chi^2$.
Only varying the scale height, this model is $0.8 \sigma$ worse compared to the ``best" model (with all parameters free) by $\Delta \chi^2 = 1.78$ and the dof difference of 2.
Similarly, the disk normalization model is $1.5\sigma$ away from the ``best" model.
Although the scale height model is preferred, it is not a large statistical difference between these two models with varied disk shapes.
Compared to models where the disk is varied, the model with CGM-only differences is less preferred since it is $2.9 \sigma$ away from the ``best" model.

In the different scale height models, the northern and the southern hemispheres have scale heights of $3.5\pm0.5\rm~kpc$ and $2.3\pm0.4\rm~kpc$, respectively.
The difference of the scale heights is about $1.2\rm~kpc$, which is at about $2\sigma$. 
As shown in Fig. \ref{NS_sh}, a larger scale height leads to a larger scatter at low latitudes, which is favored by the observations.
In this model, the variation of the model parameters does not affect the mass estimation in Section 4.2 and 4.3, which are all within $1\sigma$.
Therefore, we do not report new values for the masses of both the disk and the CGM. 

\subsection{The Possible Non-Uniform Structures}
It is well known that the warm gas disk and CGM of the MW both have lots of structures, e.g., the Fermi Bubbles (FBs), HVCs, and the Local Bubble.
These structures may have a non-uniform contribution to the measured column density of warm gases, which is opposite to our assumption that the density profile of the warm gas can be modeled by smooth functions.
Therefore, we adopt the blocking method to test whether these possible non-uniform structures affect our fittings; a similar method has also been used by \citetalias{Zheng:2019aa} to study the underlying gaseous structures in the MW halo. 
For the blocking, we mean to block some part of the sky to obtain a new set of sample and fitting result. 
The (non-)consistency between blocked and unblocked fitting results show the hints for the effect of the possible structures in the blocked region.

First, we consider a known structure -- the FBs.
{\citet{Bordoloi:2017aa} and \citet{Karim:2018aa} showed the enhancement of HVCs due to the FBs (both the northern and southern bubbles), which are not included in our modeling.}
Therefore, we block the sky region of $ -60^\circ < b < 60^\circ$ and $-30^\circ < l < 30^\circ$ to avoid the AGN sightlines (17 for \ion{Si}{4} and 5 for \ion{O}{6}) througth the FB.
The stellar sightlines are not masked out since none of the stars are distant enough to be in the FB.
The fitting results are $R_0 = 5.5\pm 1.1\rm~ kpc$, $z_0 = 2.9 \pm 0.5$ kpc, and the CGM difference of $0.80\pm0.34$ dex.
This solution is within $0.5 \sigma$ from the fiducial model (Table \ref{joint_fits}).
The mass estimates of the warm gas are all within uncertain of 0.2 dex for both the disk and the CGM.
Therefore, we suggest that the FBs do not contribute to the low-intermediate velocity warm gas significantly, although detailed studies on the FBs show evidence for the enhancement of HVCs \citep{Bordoloi:2017aa, Karim:2018aa}.

Then, we consider possible unknown large scale structures, such as the possible connection with HVCs and IVCs \citep{Sembach:2003aa}.
\citetalias{Zheng:2019aa} used the block bootstrapping to study it, while we consider this in a simple way.
Following \citetalias{Zheng:2019aa}, the sky is divided into eight regions with $90^\circ\times90^\circ$ based on the latitude and the longitude.
Each region is blocked out in turn, so we have eight new samples with about 7/8 sightlines of the fiducial sample.
We applied the joint model to these new samples.
The fittings of the blocked sample lead to the parameter region of $R_0 \approx 5.2 - 7.4$ kpc, $z_0 \approx 2.2- 3.0$ kpc, and $\Delta \log N^{\rm CGM} \approx 0.63-1.23$.
These results are all within uncertainty ($1 \sigma$) of the fiducial model, which indicates that there are no significant contributions from the unknown structures, and our assumption of the smooth profiles is roughly hold at large-scale.

\section{Summary}
We develop a 2-D disk-CGM model for the MW absorption line samples of \ion{Si}{4} and \ion{O}{6}.
The radial density profile of the disk is introduced to determine if it alleviates the tension between the stellar sample and the AGN sample, where a thick warm disk is supported by the stellar sample \citepalias{Savage:2009aa}, but not by the two-component disk-CGM model of the AGN sample \citepalias{Zheng:2019aa}.
More details can be found in Section 4.1 for the difference between the new model and the previous studies (e.g., \citetalias{Savage:2009aa, Zheng:2019aa}).
Adopting the new model, we obtain the scale heights and the scale lengths for the warm gas disk traced by \ion{Si}{4} and \ion{O}{6}, and estimate the masses in both the gaseous disk and the gaseous halo.
Here, we summarize our results:
\begin{enumerate}
\item For the MW, the preferred warm gas distribution has a 2-D disk component ($\rm R_EZ_E$) with exponential radial and vertical profiles ($n(r, z) = n_0\exp (-|z|/z_0)\exp(-r/r_0)$) and an anisotropic CGM component (depending on Galactic latitude).
The joint fitting of \ion{Si}{4} and \ion{O}{6} shows that these two ions could be modeled by the same density profile, which has a scale length of $r_0 = 6.1\pm1.2$ kpc and a scale height of $z_0 = 2.6\pm 0.4$ kpc.
The same shape of \ion{Si}{4} and \ion{O}{6} might indicate that these two ions are physically associated with each other despite a significant difference in their ionization potentials.
This scale length is larger than the \ion{H}{1} disk ($\approx 3-4$ kpc) and the stellar disk ($\approx 2$ kpc).
In \citetalias{Savage:2009aa}, \ion{O}{6} was found to have a lower scale height than \ion{Si}{4}, which was suggested as an evidence for different ionization mechanisms between \ion{Si}{4} and \ion{O}{6}.
However, our fitting shows that there is no significant difference between the \ion{Si}{4} and the \ion{O}{6} scale heights, but this does not mean that these two ions are cospatial.

\item From our best-fit model (the $\rm R_EZ_E$ disk and anisotropic CGM), the total mass of the warm gas disk ($\log T\approx5$) is about $\log (M/M_\odot)_{\rm SiIV} = (7.6 \pm 0.2) - \log  \frac{Z}{Z_\odot}$.

\item The CGM component in our model makes a comparable contribution of the column density as the warm gas disk.
Our modeling indicates that it has a higher column density in the direction perpendicular to the disk than the direction along the disk at $> 4 \sigma$ levels for both \ion{Si}{4} and \ion{O}{6}.
Combining these two ions, the difference is $0.82\pm 0.32$ dex at about $6.3 \sigma$ between the vertical and radial directions. 
However, some of this difference may be due to an interaction layer close to the disk, which we attribute to the CGM.

\item The mass of the low-intermediate velocity ($|v|\lesssim 100\kms$) warm ($\log T \approx 5$) gas in the CGM is estimated to be $\log (M/M_\odot) \approx 9.8\pm 0.2$ with a uniform density distribution and a metallicity of $0.5~Z_\odot$.
When we adopt a $\beta$-model (power law; $n(r) = n_0 r^{-3\beta}$) to approximate the density profile to 250 kpc, the total mass will be reduced to $\log (M/M_\odot) \approx 9.5$ ($\beta=1/3$), $\log (M/M_\odot) \approx 9.2$ ($\beta=1/2$), and $\log (M/M_\odot) \approx 8.9$ ($\beta=2/3$).
Then, the total mass of the warm CGM is estimated to be $\log (M/M_\odot)=9.5-9.8$ for the MW, combining with the HVC mass of $\log (M/M_\odot) = 9.4$ for the MW.

\item The projected column density ($\log N\sin |b|$) of AGN indicates a significant north-south asymmetry.
Our models suggest that this asymmetry is more likely due to an asymmetric disk rather than an asymmetric CGM at about $2\sigma$.
For the asymmetric disk, the variation of the density or the scale height cannot be distinguished, but the model with varying scale heights shows a smaller reduced $\chi^2$ (at $\approx 0.7\sigma$).
The northern and the southern hemispheres have scale heights of $3.5\pm 0.5\rm~kpc$ and $2.3\pm0.4\rm~kpc$, respectively.
\end{enumerate}

\acknowledgments
We thank the anonymous referee for her/his constructive comments and suggested improvements.
The authors also thank for Yong Zheng, Blair Savage,  Josh Peek, Mary Putman, and Jiangtao Li for their thoughtful discussions and helpful comments.
We also acknowledge the Hubble Spectroscopic Legacy Archive (HSLA) and the software Astropy, and their developers \citep{Astropy-Collaboration:2013aa, Peeples:2017aa}.
Z.Q. and J.N.B. would like to acknowledge support from NASA grants NNX15AM93G and NNX16AF23G.

\bibliographystyle{apj}
\bibliography{MissingBaryon}

\end{document}